\documentclass[journal]{IEEEtran}

\usepackage{amsmath}
\usepackage{amssymb}
\usepackage[scr=boondoxo]{mathalpha}
\usepackage{graphicx}
\usepackage{cite}

\graphicspath{{source/figures/}}

\begin{document}

\title{Sensor-Limited Observability and Carrier-Induced Reachability of Low-Order Rotor-Coupled NVH in Production Electric Drives\\[0.5ex]{\large A Magnetic Co-Energy, Gramian, and Active Projection Framework\\for Production-Signal Feasibility Analysis}}
\author{Meng-Chou~Wu%
\thanks{Independent Research, Taiwan
(e-mail: \mbox{martin.mc.wu@gmail.com}).}}
\maketitle

\begin{abstract}
This paper is motivated by a practical rotor-coupled Order-2 NVH problem observed under load in a production electric drive. The issue appears as a low-order vibration and noise component that can contribute to acoustic discomfort, durability risk, and vibration-induced wear. In high-speed traction motors, conventional mechanical mitigation is difficult because the rotor is a compact rotating structure with limited air-gap clearance, strict strength and balance requirements, and strong electromagnetic performance constraints. As a result, increasing structural stiffness or adding damping through conventional mechanical changes is often impractical after the motor architecture has been fixed.

At the same time, production electric-drive controllers typically do not measure rotor vibration or housing acceleration directly. They mainly have access to electrical and rotor-position signals such as ${I}_{d}/{I}_{q}$, voltage commands, resolver angle, and resolver speed. This creates a fundamental feasibility question: Can a mechanically important low-order rotor-coupled vibration be inferred or influenced using only production-accessible signals?

Although the motivating case is Order-2 NVH, this paper formulates the problem for selected low-order rotor-coupled deformation orders,

\[
r\in \{1,2,3,4\}
\]

because low-order components tend to produce larger mechanical amplitudes and are especially relevant to durability, acoustic radiation, ride comfort, and perceived electric-drive refinement. The selected vibration component is represented as an equivalent air-gap deformation mode with state

\[
{x}_{g}={[g,{v}_{g}]}^{T}
\]

where $g$ is the generalized air-gap deformation coordinate and ${v}_{g}$ is its modal velocity.

Starting from a magnetic co-energy formulation, this paper derives dq flux linkages, incremental inductance, current dynamics, electromagnetic torque, and generalized air-gap force. The passive linearized model is then analyzed using finite-time Gramians for current-based observability, resolver-based observability, and plant-level voltage-to-air-gap reachability. The analysis identifies the flux-linkage sensitivity vector ${\ell}_{\lambda}$ as the common physical bottleneck for passive current observability and passive voltage reachability, while resolver observability is governed by a separate torque-projection condition.

This paper further shows that a low-order air-gap deformation may be mechanically significant while remaining first-order invisible to passive dq electrical signals. This occurs because production electrical measurements observe a winding-level global electromagnetic projection rather than the local air-gap flux redistribution itself. This passive degeneracy motivates an active carrier-induced projection mechanism. By injecting a physically realizable electromagnetic carrier, the controller may create carrier-dependent measurement and force projections, denoted by ${\Gamma}_{\mathscr{y}}\left(m\right)$ and ${\Gamma}_{F}\left(m\right)$. These projections define carrier-on observability and reachability Gramians and provide measurable sideband signatures in production residuals.

The resulting framework reframes the original Order-2 NVH problem as a sensor-limited observability and carrier-induced reachability problem for low-order rotor-coupled modes. To make the projection mechanisms computationally explicit, this paper includes two physically anchored numerical feasibility studies. The first evaluates passive low-order projection under a 48-slot, 8-pole architecture-aware symmetry proxy. The second evaluates the residual detectability threshold for carrier-induced sidebands under finite-window noise and near-order disturbance leakage. These studies do not constitute FEM-calibrated motor validation, but they demonstrate the computational structure of the proposed framework. FEM-calibrated magnetic maps, carrier-on LTV/LTP simulation, controller dynamics, inverter nonidealities, and experimental validation remain future work.
\end{abstract}


\section{Introduction}
\label{sec:introduction}

This paper develops a control-oriented feasibility framework for analyzing sensor-limited observability and carrier-induced reachability of low-order rotor-coupled NVH in production electric drives \cite{soresini2024noise,wang2023modelling}. The analysis focuses on selected low-order rotor-coupled deformation orders, $r\in \{1,2,3,4\}$, where $r$ denotes the spatial order of the air-gap deformation component under consideration. The motivating production case corresponds to $r=2$, but the mathematical framework is written for a general selected low-order mode within this set.

The key modeling step is to represent the selected rotor-coupled vibration component as an equivalent air-gap deformation mode. The air-gap state is defined as

\[
{x}_{g}=\begin{bmatrix}g \\ {v}_{g}\end{bmatrix}
\]

where $g$ denotes the generalized air-gap deformation coordinate of the selected low-order mode and ${v}_{g}$ denotes its modal velocity. The corresponding spatial mode shape is written as

\[
{\phi}_{r}(\beta)=\cos(r\beta+{\phi}_{g})
\]

where $\beta$ is the rotor-fixed circumferential coordinate. This reduced representation does not attempt to reconstruct the full structural deformation field. Instead, it isolates the electromagnetic projection of the selected rotor-coupled mode that can interact with dq currents, resolver response, voltage commands, and carrier-induced electromagnetic forcing.

The electromechanical coupling is expressed through the chain

\[
g\to {\Lambda}_{g}\to {\lambda}_{d}{,\lambda}_{q}\to {I}_{d},{I}_{q},{T}_{e},{F}_{e,g}
\]

Here, the deformation coordinate $g$ modifies the air-gap permeance ${\Lambda}_{g}$, which changes dq flux linkages ${\lambda}_{d}{,\lambda}_{q}$, incremental inductance, electromagnetic torque ${T}_{e}$, and generalized air-gap force ${F}_{e,g}$. This chain provides the physical basis for evaluating whether the unmeasured air-gap state can be inferred from production electrical or resolver signals and whether it can be influenced through production-accessible electromagnetic input.

This paper addresses four feasibility questions. First, can $g$ and ${v}_{g}$ be observed from ${I}_{d}/{I}_{q}$ current residuals? Second, can $g$ and ${v}_{g}$ be observed from resolver angle and speed, ${\theta}_{r},{\omega}_{r}$? Third, can voltage commands ${V}_{d}/{V}_{q}$ influence $g$ and ${v}_{g}$ through the passive electromagnetic plant? Fourth, if the passive dq projection is degenerate, can an injected carrier create an additional measurement or force projection that makes the selected low-order mode observable or reachable through production-accessible signals?

Existing carrier-injection, fault-diagnosis, and harmonic-current methods provide important background but address different primary objectives. High-frequency carrier injection in electric drives is commonly used to extract saliency-dependent information for sensorless rotor-position or flux-angle estimation \cite{corley1998rotor,jebai2012signal,jebai2012sensorless}. Eccentricity and fault-diagnosis studies often interpret current, voltage, or sideband signatures as indicators of machine faults or abnormal operating conditions \cite{nandi2005condition,aggarwal2019review,ebrahimi2009eccentricity,ebrahimi2014advanced,goktas2016discernment,haddad2017voltage}. Active NVH and harmonic-current-injection approaches often begin after a target torque ripple, radial-force harmonic, or measurable vibration response has already been identified \cite{jahns1996pulsating,yan2019torque,erken2016online,dai2026data,soresini2025numerical}. In contrast, this paper does not formulate the problem as rotor-position estimation, fault classification, or direct harmonic cancellation. The selected rotor-coupled deformation is modeled as a dynamic air-gap state ${x}_{g}=\left[g\quad {v}_{g}\right]^{T}$, and the central question is whether this unmeasured modal state has any production-accessible measurement path or electromagnetic force path before a cancellation controller is designed.

The contribution is therefore not the use of carrier injection alone. Rather, the carrier is treated as an active electromagnetic projection mechanism that may change the observability and reachability of an otherwise passive-invisible low-order air-gap deformation state. In the passive analysis, the flux-linkage sensitivity vector ${\ell}_{\lambda}$ identifies when current residuals and plant-level voltage perturbations have a first-order coupling path to the selected mode. In the carrier-induced analysis, the projection coefficients ${\Gamma}_{\mathscr{y}}\left(m\right)$ and ${\Gamma}_{F}\left(m\right)$ describe whether a physically realizable carrier can create additional measurement or force projections. This positioning separates the present feasibility framework from conventional sensorless-control, fault-diagnosis, and active-NVH formulations.

The analysis begins with a magnetic co-energy formulation. The co-energy model defines dq flux linkages, incremental flux-current coupling, electromagnetic torque, and generalized air-gap force in an energy-consistent form. The selected air-gap deformation coordinate enters the electromagnetic model through permeance modulation, enabling the flux-linkage and inductance sensitivities with respect to $g$ to be derived explicitly.

These sensitivities are embedded into a nonlinear six-state electromechanical model,

\[
x=\begin{bmatrix}{I}_{d} & {I}_{q} & {\theta}_{r} & {\omega}_{r} & g & {v}_{g}\end{bmatrix}^{T}
\]

which is linearized around a fixed-speed, fixed-current loaded operating trajectory. The resulting perturbation model is expressed through a state transition matrix and used to define three finite-time Gramians: current observability, resolver observability, and voltage reachability.

The passive Gramian analysis identifies the flux-linkage sensitivity vector

\[
{\ell}_{\lambda}=\left[\frac{{\bar{\lambda}}_{d,g}}{{\bar{\lambda}}_{q,g}}\right]
\]

as the shared coupling quantity governing passive current observability and passive voltage reachability. Resolver observability is governed by a separate torque-projection margin. The local coupling analysis then interprets these conditions as spatial projection requirements and shows how passive degeneracy can occur when the selected low-order deformation is orthogonal to the dq electromagnetic projection.

The final technical part of this paper introduces carrier-induced electromagnetic projection. With an injected carrier, the relevant quantities become the carrier-dependent measurement projection ${\Gamma}_{\mathscr{y}}\left(m\right)$ and ${\Gamma}_{F}\left(m\right)$. These projections define carrier-on observability and reachability conditions, while order-domain sideband detection provides an implementation-level signature in production residuals.

To support the analytical framework without overstating validation, this paper further includes two physically anchored numerical feasibility studies. The first study evaluates whether the passive low-order projection collapse predicted by the ideal harmonic argument persists under a 48-slot, 8-pole architecture-aware symmetry proxy. The second study treats the composite carrier-induced projection strength as a free parameter and evaluates the residual sideband SNR required for finite-window detectability under noise and near-order disturbance leakage. These studies are not FEM-calibrated production-motor simulations. Their purpose is to demonstrate the computational structure, scaling behavior, and limitations of the proposed observability and carrier-induced projection framework.

The remainder of this paper is organized as follows. Section~\ref{sec:governing-equations} presents the control-oriented governing equations. Section~\ref{sec:magnetic-coenergy-closure} defines the magnetic co-energy closure and low-order air-gap deformation parameterization. Section~\ref{sec:linearization-loaded-trajectory} derives the linearized perturbation model around a loaded operating trajectory. Section~\ref{sec:state-transition-gramians} defines the state transition matrix and finite-time Gramians. Section~\ref{sec:passive-coupling-degeneracy} derives local passive coupling conditions and passive degeneracy mechanisms. Section~\ref{sec:carrier-induced-projection} develops the carrier-induced electromagnetic projection framework for observability and reachability. Section~\ref{sec:numerical-feasibility} presents two physically anchored numerical feasibility studies: an architecture-aware passive projection study and a carrier-induced residual detectability study. Section~\ref{sec:conclusions-validation} summarizes the conclusions and outlines a high-level future validation roadmap.

\section{Control-Oriented Governing Equations}
\label{sec:governing-equations}

This section establishes the nonlinear governing equations used throughout this paper. The objective is to define a control-oriented electromechanical model before introducing the detailed magnetic co-energy closure and air-gap deformation parameterization. The model contains dq current dynamics, rotor mechanical dynamics, and a reduced air-gap deformation mode representing the selected low-order rotor-coupled NVH component.

\subsection{State, Inputs, Disturbances, and Measurements}
\label{subsec:state-inputs-disturbances-measurements}

The state vector is defined as

\[
x=\begin{bmatrix}{I}_{d} & {I}_{q} & {\theta}_{r} & {\omega}_{r} & g & {v}_{g}\end{bmatrix}^{T}
\]

where ${I}_{d}$ and ${I}_{q}$ are dq-axis currents, ${\theta}_{r}$ is the rotor mechanical angle, ${\omega}_{r}$ is the rotor mechanical speed, $g$ is the generalized air-gap deformation coordinate of the selected low-order rotor-coupled mode, and ${v}_{g}$ is its modal velocity. Although $g$ and ${v}_{g}$ are included in the model state, they are not production measurements. They must be inferred indirectly from electrical residuals, resolver signals, or carrier-induced response signatures.

The air-gap state is

\[
{x}_{g}=\begin{bmatrix}g \\ {v}_{g}\end{bmatrix}\qquad {v}_{g}=\dot{g}
\]

The voltage command input is

\[
u=\begin{bmatrix}{V}_{d} \\ {V}_{q}\end{bmatrix}
\]

Two disturbance inputs are considered:

\[
d=\begin{bmatrix}{T}_{load} \\ {F}_{load,g}\end{bmatrix}
\]

where ${T}_{load}$ is the external load torque applied to the rotor mechanical dynamics, and ${F}_{load,g}$ is an external generalized force acting on the air-gap deformation coordinate.

The production current measurement channel is

\[
{\mathscr{y}}_{I}=\begin{bmatrix}{I}_{d} \\ {I}_{q}\end{bmatrix}={C}_{I}x\qquad {\mathscr{y}}_{R}=\begin{bmatrix}{\theta}_{r} \\ {\omega}_{r}\end{bmatrix}={C}_{R}x
\]

with

\[
{C}_{I}=\begin{bmatrix}1 & 0 & 0 & 0 & 0 & 0 \\ 0 & 1 & 0 & 0 & 0 & 0\end{bmatrix}
\]

\[
{C}_{R}=\begin{bmatrix}0 & 0 & 1 & 0 & 0 & 0 \\ 0 & 0 & 0 & 1 & 0 & 0\end{bmatrix}
\]

These two measurement channels are treated separately in the later Gramian analysis because current sensing and resolver sensing interact with the air-gap deformation state through different physical paths.

\subsection{Co-Energy-Based Electromagnetic Variables}
\label{subsec:co-energy-variables}

The electromagnetic model is written in terms of a dq-normalized magnetic co-energy function, following standard electric-machine reference-frame and energy-based modeling conventions \cite{krause2013analysis,krishnan2010permanent,jebai2014energy,jebai2011estimation},

\[
{\mathscr{w}}_{c}={\mathscr{w}}_{c}({I}_{d},{I}_{q},{\theta}_{r},g)
\]

The dq flux linkages are defined by

\[
{\lambda}_{d}=\frac{\partial {\mathscr{w}}_{c}}{\partial {I}_{d}}\qquad {\lambda}_{q}=\frac{\partial {\mathscr{w}}_{c}}{\partial {I}_{q}}
\]

The incremental flux-current matrix is

\[
{M}_{\lambda}=\frac{\partial ({\lambda}_{d},{\lambda}_{q})}{\partial ({I}_{d},{I}_{q})}=\begin{bmatrix}{\lambda}_{d,{I}_{d}} & {\lambda}_{d,{I}_{q}} \\ {\lambda}_{q,{I}_{d}} & {\lambda}_{q,{I}_{q}}\end{bmatrix}
\]

For an energy-consistent magnetic co-energy model,

\[
{\lambda}_{d,{I}_{q}}={\lambda}_{q,{I}_{d}}
\]

so ${M}_{\lambda}$ is symmetric. The current dynamics require ${M}_{\lambda}$ to be locally invertible. Therefore, the regularity condition is

\[
det\left({M}_{\lambda}\right)\neq 0
\]

A stronger physical condition is positive definiteness,

\[
{M}_{\lambda}\succ 0
\]

which corresponds to local convexity of the magnetic co-energy with respect to current.

The total three-phase electromagnetic co-energy is represented by the factor 3/2 multiplying the dq-normalized co-energy. This convention is used consistently in the electromagnetic torque and generalized force definitions below.

\subsection{\(dq\) Electrical Dynamics}
\label{subsec:dq-electrical-dynamics}

The dq voltage equations are

\[
{V}_{d}={R}_{s}{I}_{d}+{\dot{\lambda}}_{d}-p{\omega}_{r}{\lambda}_{q}
\]

\[
{V}_{q}={R}_{s}{I}_{q}+{\dot{\lambda}}_{q}+p{\omega}_{r}{\lambda}_{d}
\]

where ${R}_{s}$ is the stator phase resistance and $p$ is the pole-pair number.

Since ${\lambda}_{d}$ and ${\lambda}_{q}$ depend on ${I}_{d}$, ${I}_{q}$, ${\theta}_{r}$, and $g$, their time derivatives are

\[
{\dot{\lambda}}_{d}={\lambda}_{d,{I}_{d}}{\dot{I}}_{d}+{\lambda}_{d,{I}_{q}}{\dot{I}}_{q}+{\lambda}_{d,{\theta}_{r}}{\omega}_{r}+{\lambda}_{d,g}{v}_{g}
\]

\[
{\dot{\lambda}}_{q}={\lambda}_{q,{I}_{d}}{\dot{I}}_{d}+{\lambda}_{q,{I}_{q}}{\dot{I}}_{q}+{\lambda}_{q,{\theta}_{r}}{\omega}_{r}+{\lambda}_{q,g}{v}_{g}
\]

In matrix form,

\[
\left[\frac{{\dot{\lambda}}_{d}}{{\dot{\lambda}}_{q}}\right]={M}_{\lambda}\left[\frac{{\dot{I}}_{d}}{{\dot{I}}_{q}}\right]+\left[\frac{{\lambda}_{d,{\theta}_{r}}}{{\lambda}_{q,{\theta}_{r}}}\right]{\omega}_{r}+\left[\frac{{\lambda}_{d,g}}{{\lambda}_{q,g}}\right]{v}_{g}
\]

Solving the voltage equations for the current derivatives gives

\[
\left[\frac{{\dot{I}}_{d}}{{\dot{I}}_{q}}\right]={M}_{\lambda}^{-1}\left[\frac{{r}_{d}}{{r}_{q}}\right]
\]

where

\[
{r}_{d}={V}_{d}{-R}_{s}{I}_{d}+p{\omega}_{r}{\lambda}_{q}-{\omega}_{r}{\lambda}_{d,{\theta}_{r}}-{v}_{g}{\lambda}_{d,g}
\]

\[
{r}_{q}={V}_{q}{-R}_{s}{I}_{q}-p{\omega}_{r}{\lambda}_{d}-{\omega}_{r}{\lambda}_{q,{\theta}_{r}}-{v}_{g}{\lambda}_{q,g}
\]

These equations show how the air-gap deformation state enters the electrical dynamics through the flux-linkage maps and their derivatives with respect to $g$.

\subsection{Rotor Mechanical Dynamics}
\label{subsec:rotor-mechanical-dynamics}

The rotor mechanical dynamics are

\[
{\dot{\theta}}_{r}={\omega}_{r}
\]

\[
{\dot{\omega}}_{r}=\frac{1}{J}({T}_{e}-B{\omega}_{r}-{T}_{load})
\]

where $J$ is the rotor inertia and $B$ is the viscous damping coefficient of the rotor mechanical coordinate. The electromagnetic torque is derived from the magnetic co-energy as

\[
{T}_{e}=\frac{3}{2}[p\left({\lambda}_{d}{I}_{q}-{\lambda}_{q}{I}_{d}\right)+{\mathscr{w}}_{c,{\theta}_{r}}]
\]

The first term is the conventional dq electromagnetic torque contribution. The second term accounts for explicit mechanical-angle dependence of the co-energy, such as spatial magnetic saliency or other angle-dependent magnetic effects. Since ${\theta}_{r}$ is the mechanical rotor angle, the derivative ${\mathscr{w}}_{c,{\theta}_{r}}$ is taken with respect to mechanical angle and does not receive an additional pole-pair multiplier.

This torque expression provides the pathway by which the air-gap deformation can become visible in resolver speed. If $g$ changes the flux linkages or the angle-dependent co-energy term, then it may perturb ${T}_{e}$, which can then propagate to ${\omega}_{r}$ and ${\theta}_{r}$.

\subsection{Air-Gap Mode Dynamics}
\label{subsec:air-gap-mode-dynamics}

The air-gap deformation mode is modeled as a second-order generalized coordinate:

\[
\dot{g}={v}_{g}
\]

\[
{\dot{v}}_{g}=\frac{1}{{m}_{g}}({F}_{e,g}-{c}_{g}{v}_{g}-{k}_{g}g+{F}_{load,g})
\]

where ${m}_{g}$, ${c}_{g}$, and ${k}_{g}$ are the equivalent modal mass, damping, and stiffness of the air-gap deformation coordinate. The term ${F}_{load,g}$ represents external generalized forcing from load-path excitation, structural interaction, or other non-electromagnetic sources projected onto the air-gap mode. The coordinate $g$ is measured relative to the selected loaded operating equilibrium. Therefore, constant electromagnetic or load-path bias forces are absorbed into the equilibrium definition, while ${F}_{load,g}$ represents perturbation forcing around that operating point.

The electromagnetic generalized force is defined as

\[
{F}_{e,g}={s}_{g}\frac{3}{2}{\mathscr{w}}_{c,g}
\]

where ${s}_{g}$ is a sign convention factor that maps the positive direction of $g$ to the positive generalized electromagnetic force direction. The derivative ${\mathscr{w}}_{c,g}$ captures how the magnetic co-energy changes with the air-gap deformation coordinate. This is the main coupling through which dq currents and magnetic field energy can excite or modify the air-gap mode.

\subsection{Modeling Scope of the Air-Gap Coordinate}
\label{subsec:air-gap-coordinate-scope}

The coordinate $g$ is a reduced electromagnetic projection of the selected low-order rotor-coupled vibration component. It should not be interpreted as a complete structural description of the rotor, shaft, bearings, stator, or load path. In a more general representation, the air-gap deformation field can be expanded as

\[
\delta g\left(\beta,t\right)=\sum_{k=1}^{{N}_{g}}{q}_{k}(t){\phi}_{k}(\beta)
\]

where each generalized coordinate ${q}_{k}$ represents a different spatial deformation component. Each component would then have its own electromagnetic sensitivity vector,

\[
{\ell}_{\lambda,k}=\left[\frac{{\bar{\lambda}}_{d,{q}_{k}}}{{\bar{\lambda}}_{q,{q}_{k}}}\right]
\]

The present single-coordinate model corresponds to one selected low-order rotor-coupled deformation component. The motivating production case is $r=2$, but the formulation is written for a general selected order $r\in \{1,2,3,4\}$. This reduction is used to expose the core coupling mechanism between air-gap deformation, dq flux-linkage sensitivity, current residuals, resolver response, and voltage-to-air-gap reachability.

The single-coordinate model should therefore be interpreted as a minimal control-oriented electromagnetic projection, not as a complete structural reduction of the rotor-bearing-stator system. Its purpose is to reveal the dominant projection and degeneracy mechanisms under production-signal constraints. Additional radial, tangential, bending, tilting, load-path, or three-dimensional deformation coordinates can be appended when FEM or experimental identification indicates that they contribute significantly to the measured response.

\subsection{Compact Nonlinear State-Space Form}
\label{subsec:compact-nonlinear-state-space}

Combining the current dynamics, rotor dynamics, and air-gap dynamics gives the nonlinear state-space model

\[
\dot{x}=f(x,u,d)
\]

where

\[
x=\begin{bmatrix}{I}_{d} \\ {I}_{q} \\ {\theta}_{r} \\ {\omega}_{r} \\ g \\ {v}_{g}\end{bmatrix}\qquad u=\begin{bmatrix}{V}_{d} \\ {V}_{q}\end{bmatrix}\qquad d=\begin{bmatrix}{T}_{load} \\ {F}_{load,g}\end{bmatrix}
\]

Explicitly,

\[
\dot{x}=\begin{bmatrix}{\dot{I}}_{d} \\ {\dot{I}}_{q} \\ {\dot{\theta}}_{r} \\ {\dot{\omega}}_{r} \\ \dot{g} \\ {\dot{v}}_{g}\end{bmatrix}=\begin{bmatrix}{\left[{M}_{\lambda}^{-1}r\right]}_{1} \\ {\left[{M}_{\lambda}^{-1}r\right]}_{2} \\ {\omega}_{r} \\ {1}/{J}({T}_{e}-B{\omega}_{r}-{T}_{load}) \\ {v}_{g} \\ {1}/{{m}_{g}}({F}_{e,g}-{c}_{g}{v}_{g}-{k}_{g}g+{F}_{load,g})\end{bmatrix}
\]

with

\[
r=\begin{bmatrix}{r}_{d} \\ {r}_{q}\end{bmatrix}
\]

This governing model is written before specifying the detailed magnetic co-energy closure. The next section defines the co-energy structure, introduces the low-order air-gap deformation parameterization, and derives the magnetic sensitivities that determine the observability and reachability coupling terms used later in the Gramian analysis.

\section{Magnetic Co-Energy Closure and Low-Order Air-Gap Deformation Parameterization}
\label{sec:magnetic-coenergy-closure}

This section specifies the magnetic co-energy closure used in the governing equations introduced in Section~\ref{sec:governing-equations}. The objective is to express the dq flux linkages, incremental flux-current matrix, and air-gap deformation sensitivities in terms of a physically interpretable electromagnetic model. The selected air-gap deformation coordinate $g$ enters the electromagnetic model through local permeance modulation, which modifies PM flux linkage, dq inductance, electromagnetic torque, and generalized air-gap force.

\subsection{Magnetic Co-Energy Expansion}
\label{subsec:magnetic-coenergy-expansion}

The dq-normalized magnetic co-energy is written as

\[
\begin{aligned}\mathscr{w}_{c}(I_d,I_q,\theta_r,g)={}&\mathscr{w}_{0}(\theta_r,g)+\psi_d(\theta_r,g)I_d+\psi_q(\theta_r,g)I_q \\&+\frac{1}{2}L_{dd}(\theta_r,g)I_d^2+L_{dq}(\theta_r,g)I_dI_q \\&+\frac{1}{2}L_{qq}(\theta_r,g)I_q^2+\mathscr{w}_{sat}(I_d,I_q)\end{aligned}
\]

Here, ${\mathscr{w}}_{0}\left({\theta}_{r},g\right)$ represents the current-independent magnetic co-energy contribution, including PM-field energy terms. The functions ${\psi}_{d}\left({\theta}_{r},g\right)$ and ${\psi}_{q}\left({\theta}_{r},g\right)$ represent PM-induced dq flux-linkage components. The terms ${L}_{dd}\left({\theta}_{r},g\right)$, ${L}_{dq}\left({\theta}_{r},g\right)$, and ${L}_{qq}\left({\theta}_{r},g\right)$ represent the dq inductance map, including possible saliency and air-gap deformation dependence.

The term ${\mathscr{w}}_{sat}({I}_{d}{,I}_{q})$ is a local current-saturation correction. Energy-based and saturation-aware PMSM models have been used to preserve consistency between flux linkage, incremental inductance, torque, and injected-signal response \cite{jebai2014energy,jebai2011estimation}. In this paper, it is modeled by a third-order polynomial,

\[
\begin{aligned}
\mathscr{w}_{sat}(I_d,I_q)&=\frac{1}{6}C_{ddd}I_d^3+\frac{1}{2}C_{ddq}I_d^2I_q \\
&\quad+\frac{1}{2}C_{dqq}I_dI_q^2+\frac{1}{6}C_{qqq}I_q^3
\end{aligned}
\]

The coefficients ${C}_{ddd}$, ${C}_{ddq}$, ${C}_{dqq}$, and ${C}_{qqq}$ are local phenomenological saturation coefficients. They are treated as constants in the baseline model. This approximation preserves an energy-consistent relationship between flux linkage and incremental flux-current coupling while avoiding the need for a full nonlinear saturation map at this stage.

The co-energy closure should therefore be interpreted as a local operating-point model, not as a high-fidelity magnetic replacement for FEM. Its purpose is to preserve the derivative structure needed to define flux linkages, incremental inductance, electromagnetic torque, and generalized air-gap force in a mutually consistent way. In a numerical implementation, the polynomial saturation term and the maps ${\psi}_{i}({\theta}_{r},g)$ and ${L}_{i,j}({\theta}_{r},g)$ may be replaced by FEM-calibrated or experimentally identified co-energy maps without changing the state-space and Gramian framework.

More detailed current-, air-gap-, or carrier-dependent saturation models can later be incorporated using FEM-derived co-energy data or experimental calibration.

\subsection{Flux Linkages and Incremental Flux-Current Matrix}
\label{subsec:flux-linkages-incremental-matrix}

The dq flux linkages are obtained from the co-energy derivatives,

\[
{\lambda}_{d}=\frac{\partial {\mathscr{w}}_{c}}{\partial {I}_{d}}\qquad {\lambda}_{q}=\frac{\partial {\mathscr{w}}_{c}}{\partial {I}_{q}}
\]

Using the co-energy expansion above,

\[
{\lambda}_{d}={\psi}_{d}+{L}_{dd}{I}_{d}+{L}_{dq}{I}_{q}+\frac{1}{2}{C}_{ddd}{I}_{d}^{2}+{C}_{ddq}{I}_{d}{I}_{q}+\frac{1}{2}{C}_{dqq}{I}_{q}^{2}
\]

\[
{\lambda}_{q}={\psi}_{q}+{L}_{dq}{I}_{d}+{L}_{qq}{I}_{q}+\frac{1}{2}{C}_{ddq}{I}_{d}^{2}+{C}_{dqq}{I}_{d}{I}_{q}+\frac{1}{2}{C}_{qqq}{I}_{q}^{2}
\]

For compactness, the dependence of ${\psi}_{d}$, ${\psi}_{q}$, ${L}_{dd}$, ${L}_{dq}$, ${L}_{qq}$ on ${\theta}_{r}$ and $g$ is omitted in the equations above, but it is retained in the model. Therefore,

\[
{\lambda}_{d}={\lambda}_{d}({I}_{d},{I}_{q},{\theta}_{r},g)\qquad {\lambda}_{q}={\lambda}_{q}({I}_{d},{I}_{q},{\theta}_{r},g)
\]

The incremental flux-current matrix is defined as

\[
{M}_{\lambda}=\frac{\partial ({\lambda}_{d},{\lambda}_{q})}{\partial ({I}_{d},{I}_{q})}=\begin{bmatrix}{M}_{11} & {M}_{12} \\ {M}_{12} & {M}_{22}\end{bmatrix}
\]

where

\[
{M}_{11}={L}_{dd}+{C}_{ddd}{I}_{d}+{C}_{ddq}{I}_{q}
\]

\[
{M}_{12}={L}_{dq}+{C}_{ddq}{I}_{d}+{C}_{dqq}{I}_{q}
\]

\[
{M}_{22}={L}_{qq}+{C}_{dqq}{I}_{d}+{C}_{qqq}{I}_{q}
\]

Since ${M}_{\lambda}$ is the Hessian of the magnetic co-energy with respect to current,

\[
{M}_{\lambda}={\nabla}_{I}^{2}{\mathscr{w}}_{c}
\]

it is symmetric by construction. The current dynamics require local invertibility,

\[
\Delta=M_{11}M_{22}-M_{12}^{2}\neq 0
\]

A physically stronger condition is

\[
{M}_{\lambda}\succ 0
\]

or equivalently,

\[
M_{11}>0\qquad \Delta>0
\]

This condition corresponds to local convexity of the magnetic co-energy with respect to ${I}_{d}$ and ${I}_{q}$. Because the cubic saturation model is local, positive definiteness should be interpreted as an operating-point condition rather than a global property over all possible currents.

\subsection{Low-Order Air-Gap Deformation Parameterization}
\label{subsec:low-order-air-gap-parameterization}

The air-gap deformation coordinate $g$ is now connected to the electromagnetic model through a rotor-fixed spatial parameterization. Let $\alpha$ denote the stator-fixed mechanical circumferential coordinate, and let ${\theta}_{r}$ denote the rotor mechanical angle. The rotor-fixed coordinate is

\[
\beta=\alpha-{\theta}_{r}
\]

The selected low-order air-gap deformation component is represented by the normalized mode shape

\[
{\phi}_{r}\left(\beta\right)=\cos(r\beta+{\phi}_{g})
\]

where ${\phi}_{g}$ is the spatial phase of the projected air-gap deformation component. The effective local air gap is modeled as

\[
{g}_{eff}\left(\beta,t\right)={g}_{0}+g(t){\phi}_{r}\left(\beta\right)
\]

or

\[
{g}_{eff}\left(\beta,t\right)={g}_{0}+g(t)\cos(r\beta+{\phi}_{g})
\]

Here, ${g}_{0}$ is the nominal air-gap length, while $g(t)$ is the generalized deformation coordinate measured relative to the selected loaded operating equilibrium. The model requires the local air gap to remain positive,

\[
{g}_{0}+g\left(t\right){\phi}_{r}\left(\beta\right)>0
\]

for all relevant $\beta$. A sufficient local condition is

\[
\left|g(t)\right|\left\Vert\phi_r\right\Vert_{\infty}<g_0
\]

With the normalization $\left\Vert\phi_r\right\Vert_{\infty}=1$, this reduces to

\[
\left|g(t)\right|<g_0
\]

This parameterization should not be interpreted as a complete structural mode shape of the rotor. It is the electromagnetic air-gap projection of the selected low-order rotor-coupled vibration component.

\subsection{Air-Gap Permeance Model and Local Derivatives}
\label{subsec:air-gap-permeance-derivatives}

The local air-gap permeance is modeled as

\[
{\Lambda}_{g}(\beta,g)=\frac{{\mu}_{0}}{{g}_{0}+g{\phi}_{r}(\beta)}
\]

where ${\mu}_{0}$ is the permeability of free space. The first derivative with respect to $g$ is

\[
\frac{\partial {\Lambda}_{g}}{\partial g}=-\frac{{\mu}_{0}{\phi}_{r}(\beta)}{{\left[{g}_{0}+g{\phi}_{r}(\beta)\right]}^{2}}
\]

Evaluated at the nominal deformation coordinate $g=0$,

\[
{\left.\frac{\partial {\Lambda}_{g}}{\partial g}\right\vert}_{g=0}=-\frac{{\mu}_{0}}{{g}_{0}^{2}}{\phi}_{r}(\beta)
\]

The second derivative is

\[
\frac{{\partial}^{2}{\Lambda}_{g}}{\partial {g}^{2}}=\frac{2{\mu}_{0}{{\phi}^{2}}_{r}(\beta)}{{\left[{g}_{0}+g{\phi}_{r}(\beta)\right]}^{3}}
\]

At $g=0$,

\[
{\left.\frac{{\partial}^{2}{\Lambda}_{g}}{\partial {g}^{2}}\right\vert}_{g=0}=\frac{{2\mu}_{0}}{{g}_{0}^{3}}{{\phi}^{2}}_{r}(\beta)
\]

These permeance derivatives provide the physical source of the flux-linkage sensitivity, inductance sensitivity, and electromagnetic stiffness terms that appear later in the linearized model.

\subsection{Flux-Linkage and Inductance Maps}
\label{subsec:flux-linkage-inductance-maps}

The PM-induced dq flux-linkage components are modeled as projections of the PM magnetomotive force onto the dq winding basis through the air-gap permeance. Let ${b}_{d}(\beta)$ and ${b}_{q}(\beta)$ denote the dq winding basis functions, and let ${\mathscr{F}}_{m}(\beta)$ denote the rotor-fixed PM magnetomotive force distribution. In the baseline rotor-fixed analytical closure, the flux-linkage and inductance maps are written as functions of $g$. More general maps may retain explicit ${\theta}_{r}$-dependence if stator-fixed asymmetry, slotting interaction, or manufacturing variation is included. The PM-induced dq flux-linkage map is written as

\[
{\psi}_{i}\left(g\right)={K}_{\psi}{\int}_{0}^{2\pi}{b}_{i}(\beta){\mathscr{F}}_{m}(\beta){\Lambda}_{g}(\beta,g)d\beta
\]

\[
i\in \left\{d,q\right\}
\]

Similarly, the dq inductance map is written as

\[
{L}_{ij}\left(g\right)={L}_{ij}^{\sigma}+{K}_{L}{\int}_{0}^{2\pi}{b}_{i}(\beta){b}_{j}(\beta){\Lambda}_{g}(\beta,g)d\beta
\]

\[
i,j\in \left\{d,q\right\}
\]

Here, ${K}_{\psi}$ and ${K}_{L}$ are scaling constants that absorb winding turns, stack length, geometric factors, and normalization conventions. The terms ${L}_{ij}^{\sigma}$ represent leakage or non-air-gap inductance contributions.

The baseline dq basis can be written as

\[
{b}_{d}\left(\beta\right)=\cos(p\beta)\qquad {b}_{q}\left(\beta\right)=\sin(p\beta)
\]

where $p$ is the pole-pair number. The PM MMF can be represented by the harmonic expansion

\[
{\mathscr{F}}_{m}\left(\beta\right)={\sum}_{n\in \mathscr{H}_m}{\mathscr{F}}_{mn}\cos\left(np\beta+{\varphi}_{mn}\right)
\]

Here, $n\in\mathscr{H}_m$ denotes the PM MMF spatial-harmonic index.

The dq projection and harmonic representation are consistent with standard PMSM reference-frame and electromagnetic modeling treatments \cite{krause2013analysis,krishnan2010permanent}. These expressions are not used to claim that the simplified harmonic model fully captures the motor electromagnetic field. Instead, they provide an analytical closure for identifying which spatial components of the air-gap deformation project into dq flux-linkage and inductance sensitivities.

The resulting harmonic-selection arguments should be interpreted as symmetry-limit predictions. They identify when a selected low-order deformation is expected to be first-order invisible under an idealized full-circumference dq projection. Real machines may break this ideal symmetry through slotting, saturation, winding distribution, rotor or stator asymmetry, eccentricity, manufacturing variation, local magnetic saturation, end effects, or control-dependent carrier fields. Detailed electromagnetic NVH studies commonly rely on electromagnetic force, flux, and structural-response models to capture these nonideal effects \cite{soresini2024noise,wang2023modelling,yang2020radial}. These nonidealities can generate small but nonzero effective projection coefficients even when the ideal harmonic closure predicts cancellation.

Therefore, the analytical maps in this section should be viewed as a baseline mechanism for understanding passive degeneracy. In a later numerical implementation, ${\psi}_{i}\left(g\right)$ and ${L}_{ij}\left(g\right)$ can be obtained directly from FEM or calibrated experimental maps.

\subsection{Air-Gap Sensitivities of Flux Linkage and Inductance}
\label{subsec:air-gap-sensitivities}

Differentiating the PM flux-linkage map with respect to $g$, and evaluating at $g=0$, gives

\[
{\bar{\psi}}_{i,g}=-\frac{{K}_{\psi}{\mu}_{0}}{{g}_{0}^{2}}{\int}_{0}^{2\pi}{b}_{i}(\beta){\mathscr{F}}_{m}(\beta){\phi}_{r}(\beta)d\beta
\]

\[
i\in \left\{d,q\right\}
\]

and

\[
{\bar{L}}_{ij,g}=-\frac{{K}_{L}{\mu}_{0}}{{g}_{0}^{2}}{\int}_{0}^{2\pi}{b}_{i}(\beta){b}_{j}(\beta){\phi}_{r}(\beta)d\beta
\]

\[
i\in \left\{d,q\right\}
\]

The second derivatives are

\[
{\bar{\psi}}_{i,gg}=\frac{2{K}_{\psi}{\mu}_{0}}{{g}_{0}^{3}}{\int}_{0}^{2\pi}{b}_{i}(\beta){\mathscr{F}}_{m}(\beta){{\phi}^{2}}_{r}(\beta)d\beta
\]

\[
{\bar{L}}_{ij,gg}=\frac{2{K}_{L}{\mu}_{0}}{{g}_{0}^{3}}{\int}_{0}^{2\pi}{b}_{i}(\beta){b}_{j}(\beta){{\phi}^{2}}_{r}(\beta)d\beta
\]

These first- and second-order sensitivities determine how the air-gap deformation enters the current dynamics, electromagnetic torque, generalized air-gap force, and effective electromagnetic stiffness.

At the frozen-time nominal loaded operating point,

\[
\bar{x}=\begin{bmatrix}I_{d0} & I_{q0} & \bar{\theta}_r & \Omega & 0 & 0\end{bmatrix}^{T}
\]

the nominal flux-linkage sensitivity vector is defined as

\[
{\ell}_{\lambda}=\left[\frac{{\bar{\lambda}}_{d,g}}{{\bar{\lambda}}_{q,g}}\right]
\]

Using the flux-linkage expressions,

\[
{\bar{\lambda}}_{d,g}={\bar{\psi}}_{d,g}+{\bar{L}}_{dd,g}{I}_{d0}+{\bar{L}}_{dq,g}{I}_{q0}
\]

\[
{\bar{\lambda}}_{q,g}={\bar{\psi}}_{q,g}+{\bar{L}}_{dq,g}{I}_{d0}+{\bar{L}}_{qq,g}{I}_{q0}
\]

Thus,

\[
{\ell}_{\lambda}=\begin{bmatrix}{\bar{\psi}}_{d,g}+{\bar{L}}_{dd,g}{I}_{d0}+{\bar{L}}_{dq,g}{I}_{q0} \\ {\bar{\psi}}_{q,g}+{\bar{L}}_{dq,g}{I}_{d0}+{\bar{L}}_{qq,g}{I}_{q0}\end{bmatrix}
\]

This vector is the central electromagnetic coupling quantity of this paper. It measures how the air-gap deformation coordinate changes the dq flux linkages at the selected loaded operating point. In later sections, ${\ell}_{\lambda}$ appears in both the current-based observability path and the voltage-to-air-gap reachability path.

However, the existence of ${\ell}_{\lambda}\neq 0$ should be interpreted as a first-order coupling condition, not as a complete guarantee of practical observability or usable actuation authority under noise, controller dynamics, inverter nonidealities, voltage saturation, or finite measurement bandwidth. Conversely, a small value of ${\ell}_{\lambda}$ under the ideal analytical closure should not be interpreted as proof that all production signatures vanish in the real motor. It indicates that passive dq projection is weak or degenerate at first order and that any usable coupling must come from nonideal asymmetries, calibrated magnetic effects, load-path interaction, or active carrier-induced projection.

\subsection{Baseline Symmetry Interpretation}
\label{subsec:baseline-symmetry-interpretation}

The formulation above retains general dependence on ${\theta}_{r}$ and $g$ at the governing-equation level. This is intentional: explicit angle dependence allows the model to include spatial saliency, rotor-fixed asymmetry, manufacturing variation, or other angle-dependent magnetic effects.

For the baseline analytical closure, however, the air-gap deformation and magnetic quantities are expressed in the rotor-fixed coordinate $\beta=\alpha-{\theta}_{r}$. Under an ideal symmetric machine approximation without a low-order stator-fixed asymmetry, the resulting PM flux-linkage and inductance maps can be treated as functions of $g$ without explicit dependence on ${\theta}_{r}$. In that case,

\[
{\psi}_{i}={\psi}_{i}\left(g\right)\qquad {L}_{ij}={L}_{ij}\left(g\right)
\]

and the mixed derivatives involving explicit mechanical angle dependence may vanish,

\[
{\lambda}_{i,{\theta}_{r},g}=0\qquad {\mathscr{w}}_{c,{\theta}_{r},g}=0
\]

unless additional low-order asymmetry, geometric imperfection, or stator-fixed spatial modulation is introduced.

This baseline symmetry interpretation should not be confused with the absence of rotor-coupled NVH. The air-gap deformation coordinate $g$ still exists and can still affect dq flux linkages through permeance modulation. The statement is only that, under the baseline symmetry assumptions, explicit ${\theta}_{r}$-dependent magnetic coupling terms are not the primary mechanism. More detailed FEM-based maps can later restore angle-dependent coupling if the physical motor shows significant asymmetry, slotting interaction, dynamic eccentricity, or manufacturing-induced spatial variation.

\section{Linearization Around a Loaded Operating Trajectory}
\label{sec:linearization-loaded-trajectory}

This section linearizes the nonlinear governing equations around a fixed-speed, fixed-current loaded operating trajectory. The objective is to obtain a perturbation model that separates the current states, resolver states, and air-gap deformation states. This block structure is later used to define finite-time observability and reachability Gramians and to identify the local coupling mechanisms.

The linearization in this section should be interpreted as a local passive plant model around a representative operating window, not as a global model of the full electric-drive system. It is intended to expose the first-order electromechanical coupling paths among dq currents, rotor motion, and the selected air-gap deformation coordinate. Controller dynamics, PWM implementation, inverter saturation, current-loop rejection, thermal limits, and carrier-induced modulation are not included in this passive linearization. These effects enter later either as practical weighting terms in the finite-time Gramians or as carrier-on LTV/LTP dynamics in Section~\ref{sec:carrier-induced-projection}.

\subsection{Nominal Loaded Operating Trajectory}
\label{subsec:nominal-loaded-trajectory}

The nominal trajectory is selected as a fixed-speed, fixed-current loaded operating condition,

\[
\bar{x}(t)=\begin{bmatrix}I_{d0} \\ I_{q0} \\ \theta_0+\Omega(t-t_0) \\ \Omega \\ 0 \\ 0\end{bmatrix}
\]

where ${I}_{d0}$ and ${I}_{q0}$ are the nominal dq currents, $\Omega$ is the nominal rotor mechanical speed, and ${\theta}_{0}$ is the rotor mechanical angle at the initial time ${t}_{0}$. The air-gap deformation coordinate is measured relative to the selected loaded operating equilibrium, so the nominal deformation state is

\[
{\bar{x}}_{g}=\begin{bmatrix}0 \\ 0\end{bmatrix}
\]

The nominal voltage command $\bar{u}(t)$ is chosen such that the dq current derivatives vanish along the nominal trajectory,

\[
{\dot{\bar{I}}}_{d}=0\qquad {\dot{\bar{I}}}_{q}=0
\]

Using the dq voltage equations, the nominal voltage commands satisfy

\[
{\bar{V}}_{d}={R}_{s}{I}_{d0}-p\Omega{\bar{\lambda}}_{q}+\Omega{\bar{\lambda}}_{d,{\theta}_{r}}
\]

\[
{\bar{V}}_{q}={R}_{s}{I}_{q0}+p\Omega{\bar{\lambda}}_{d}+\Omega{\bar{\lambda}}_{q,{\theta}_{r}}
\]

where all flux linkages and derivatives are evaluated along $\bar{x}(t)$. Under the baseline symmetry assumptions in which explicit ${\theta}_{r}$-dependence is absent, these reduce to

\[
{\bar{V}}_{d}={R}_{s}{I}_{d0}-p\Omega{\bar{\lambda}}_{q}
\]

\[
{\bar{V}}_{q}={R}_{s}{I}_{q0}+p\Omega{\bar{\lambda}}_{d}
\]

The nominal load torque satisfies the rotational equilibrium condition

\[
{\bar{T}}_{load}(t)={\bar{T}}_{e}(t)-B\Omega
\]

Similarly, the nominal generalized air-gap force balance is absorbed into the definition of the loaded equilibrium for $g$. Therefore, the perturbation dynamics describe deviations around the selected loaded operating condition rather than absolute static deformation.

\subsection{Perturbation Variables}
\label{subsec:perturbation-variables}

Perturbation variables are defined as

\[
\delta x=x-\bar{x},\quad \delta u=u-\bar{u},\quad \delta d=d-\bar{d}
\]

The state perturbation is

\[
\delta x=\begin{bmatrix}\delta I_d & \delta I_q & \delta\theta_r & \delta\omega_r & \delta g & \delta v_g\end{bmatrix}^{T}
\]

The input perturbation is

\[
\delta u=\begin{bmatrix}\delta V_d \\ \delta V_q\end{bmatrix}
\]

and the disturbance perturbation is

\[
\delta d=\begin{bmatrix}\delta T_{load} \\ \delta F_{load,g}\end{bmatrix}
\]

The linearized perturbation model takes the form

\[
\delta\dot{x}=A(t)\delta x+B_u(t)\delta u+E(t)\delta d
\]

where

\[
A\left(t\right)={\left.\frac{\partial f}{\partial x}\right\vert}_{\bar{x}\left(t\right),\bar{u}\left(t\right),\bar{d}\left(t\right)}\qquad {B}_{u}\left(t\right)={\left.\frac{\partial f}{\partial u}\right\vert}_{\bar{x}\left(t\right),\bar{u}\left(t\right),\bar{d}\left(t\right)}
\]

and

\[
E\left(t\right)={\left.\frac{\partial f}{\partial d}\right\vert}_{\bar{x}\left(t\right),\bar{u}\left(t\right),\bar{d}\left(t\right)}
\]

The resulting perturbation model follows the standard local linearization form used in finite-dimensional state-space analysis \cite{rugh1996linear,kailath1980linear}. These Jacobians describe the passive plant at the selected loaded operating trajectory. If the electromagnetic maps retain explicit dependence on ${\theta}_{r}$, then $A(t)$ may be time-varying along the rotating nominal trajectory. Under the baseline rotor-fixed symmetric closure, the same formulation reduces locally to a frozen-time or constant-coefficient perturbation model around the selected operating point.

The nonlinear current dynamics can be written compactly as

\[
{\dot{x}}_{I}={M}_{\lambda}^{-1}r\qquad {x}_{I}=\begin{bmatrix}{I}_{d} \\ {I}_{q}\end{bmatrix}\qquad r=\begin{bmatrix}{r}_{d} \\ {r}_{q}\end{bmatrix}
\]

Linearizing around the nominal trajectory gives

\[
\delta\dot{x}_{I}
=
\delta\!\left(M_{\lambda}^{-1}r\right)
\]

Using the first-order variation,

\[
\delta(M_{\lambda}^{-1}r)=\bar{M}_{\lambda}^{-1}\delta r-\bar{M}_{\lambda}^{-1}(\delta M_{\lambda})\bar{M}_{\lambda}^{-1}\bar{r}.
\]

Along the fixed-current nominal trajectory,

\[
{\dot{\bar{x}}}_{I}=0\qquad \bar{r}=0
\]

Therefore, the current perturbation dynamics simplify to

\[
\delta\dot{x}_{I}=\bar{M}_{\lambda}^{-1}\delta r
\]

Equivalently, for each state component ${x}_{j}$,

\[
{A}_{I,j}\left(t\right)={\bar{M}}_{\lambda}^{-1}{\left.\frac{\partial r}{\partial {x}_{j}}\right\vert}_{\bar{x}\left(t\right),\bar{u}\left(t\right),\bar{d}\left(t\right)}
\]

The voltage input matrix for the current dynamics is

\[
{B}_{I}(t)={\bar{M}}_{\lambda}^{-1}
\]

Writing

\[
{\bar{M}}_{\lambda}=\begin{bmatrix}{\bar{M}}_{11} & {\bar{M}}_{12} \\ {\bar{M}}_{12} & {\bar{M}}_{22}\end{bmatrix}
\]

with

\[
\bar{\Delta}=\bar{M}_{11}\bar{M}_{22}-\bar{M}_{12}^{2}
\]

the inverse is

\[
\bar{M}_{\lambda}^{-1}=\frac{1}{\bar{\Delta}}\begin{bmatrix}\bar{M}_{22} & -\bar{M}_{12} \\ -\bar{M}_{12} & \bar{M}_{11}\end{bmatrix}
\]

This term maps dq voltage perturbations into dq current perturbations at the plant level.

\subsection{Linearized Rotor Mechanical Dynamics}
\label{subsec:linearized-rotor-dynamics}

The rotor mechanical equations are

\[
{\dot{\theta}}_{r}={\omega}_{r}
\]

\[
{\dot{\omega}}_{r}=\frac{1}{J}({T}_{e}-B{\omega}_{r}-{T}_{load})
\]

Linearization gives

\[
\delta{\dot{\theta}}_{r}=\delta{\omega}_{r}
\]

\[
\delta{\dot{\omega}}_{r}=\frac{1}{J}(\delta{T}_{e}-B\delta{\omega}_{r}-\delta{T}_{load})
\]

The torque perturbation can be written as

\[
\begin{aligned}
\delta T_e={}&\bar{T}_{e,I_d}\delta I_d+\bar{T}_{e,I_q}\delta I_q+\bar{T}_{e,\theta_r}\delta\theta_r \\
&\quad+\bar{T}_{e,\omega_r}\delta\omega_r+\bar{T}_{e,g}\delta g+\bar{T}_{e,v_g}\delta v_g
\end{aligned}
\]

where direct dependence of ${T}_{e}$ on ${v}_{g}$ and ${\omega}_{r}$ is neglected in the baseline co-energy model. Therefore,

\[
\begin{aligned}
\delta\dot{\omega}_r={}&\frac{1}{J}\bigl(\bar{T}_{e,I_d}\delta I_d+\bar{T}_{e,I_q}\delta I_q+\bar{T}_{e,\theta_r}\delta\theta_r \\
&\quad+\bar{T}_{e,g}\delta g-B\delta\omega_r-\delta T_{load}\bigr)
\end{aligned}
\]

The term ${\bar{T}}_{e,g}$ is the torque-projection pathway from air-gap deformation to resolver speed and angle. Its physical condition is analyzed later in the local coupling and degeneracy section.

At this stage, the resolver pathway should be interpreted as a structural coupling path in the linearized plant. Its practical conditioning depends on rotor inertia, speed-estimator bandwidth, resolver quantization, and measurement noise, which are not included in the local Jacobian itself but are accounted for later through weighted observability measures and scope limitations.

\subsection{Linearized Air-Gap Deformation Dynamics}
\label{subsec:linearized-air-gap-dynamics}

The air-gap deformation dynamics are

\[
\dot{g}={v}_{g}
\]

\[
{\dot{v}}_{g}=\frac{1}{{m}_{g}}({F}_{e,g}-{c}_{g}{v}_{g}-{k}_{g}g+{F}_{load,g})
\]

Linearization gives

\[
\delta\dot{g}=\delta{v}_{g}
\]

\[
\delta\dot{v}_g=\frac{1}{m_g}\bigl(\delta F_{e,g}-c_g\delta v_g-k_g\delta g+\delta F_{load,g}\bigr)
\]

The electromagnetic generalized-force perturbation is

\[
\begin{aligned}
\delta F_{e,g}={}&\bar{F}_{e,g,I_d}\delta I_d+\bar{F}_{e,g,I_q}\delta I_q+\bar{F}_{e,g,\theta_r}\delta\theta_r \\
&\quad+\bar{F}_{e,g,\omega_r}\delta\omega_r+\bar{F}_{e,g,g}\delta g+\bar{F}_{e,g,v_g}\delta v_g
\end{aligned}
\]

Since

\[
{F}_{e,g}={s}_{g}\frac{3}{2}{\mathscr{w}}_{c,g}
\]

the force derivatives are

\[
{F}_{e,g,{I}_{d}}={s}_{g}\frac{3}{2}{\lambda}_{d,g}\qquad {F}_{e,g,{I}_{q}}={s}_{g}\frac{3}{2}{\lambda}_{q,g}
\]

\[
{F}_{e,g,{\theta}_{r}}={s}_{g}\frac{3}{2}{\mathscr{w}}_{c,g{\theta}_{r}}\qquad {F}_{e,g,g}={s}_{g}\frac{3}{2}{\mathscr{w}}_{c,gg}
\]

The baseline model does not include direct dependence of ${F}_{e,g}$ on ${\omega}_{r}$ or ${v}_{g}$, so

\[
{F}_{e,g,{\omega}_{r}}=0\qquad {F}_{e,g,{v}_{g}}=0
\]

Therefore,

\[
\begin{aligned}\delta\dot{v}_g=\frac{1}{m_g}\Bigl[{}&s_g\frac{3}{2}\bar{\lambda}_{d,g}\delta I_d+s_g\frac{3}{2}\bar{\lambda}_{q,g}\delta I_q \\&+s_g\frac{3}{2}\bar{\mathscr{w}}_{c,g\theta_r}\delta\theta_r+\left(s_g\frac{3}{2}\bar{\mathscr{w}}_{c,gg}-k_g\right)\delta g \\&-c_g\delta v_g+\delta F_{load,g}\Bigr]\end{aligned}
\]

This equation shows that the same flux-linkage sensitivity components ${\bar{\lambda}}_{d,g}$ and ${\bar{\lambda}}_{q,g}$ govern the current-to-air-gap force pathway. Non-electromagnetic load-path forcing, structural cross-coupling, bearing interaction, and other unmodeled disturbances are represented at this level by $\delta{F}_{load,g}$. If future FEM or experimental identification shows that additional deformation coordinates or nonconservative force terms are significant, they can be appended to the air-gap subsystem without changing the basic observability and reachability construction.

\subsection{Block State Partition}
\label{subsec:block-state-partition}

For the observability and reachability analysis, the perturbation state is partitioned into three two-dimensional blocks:

\[
\delta x_I=\begin{bmatrix}\delta I_d \\ \delta I_q\end{bmatrix}\qquad \delta x_R=\begin{bmatrix}\delta\theta_r \\ \delta\omega_r\end{bmatrix}\qquad \delta x_g=\begin{bmatrix}\delta g \\ \delta v_g\end{bmatrix}
\]

Thus,

\[
\delta x=\begin{bmatrix}\delta x_I \\ \delta x_R \\ \delta x_g\end{bmatrix}
\]

The linearized model can then be written as

\[
\begin{aligned}\begin{bmatrix}\delta\dot{x}_I \\ \delta\dot{x}_R \\ \delta\dot{x}_g\end{bmatrix}={}&\begin{bmatrix}A_{II} & A_{IR} & A_{Ig} \\ A_{RI} & A_{RR} & A_{Rg} \\ A_{gI} & A_{gR} & A_{gg}\end{bmatrix}\begin{bmatrix}\delta x_I \\ \delta x_R \\ \delta x_g\end{bmatrix}\\&+\begin{bmatrix}B_I \\ 0 \\ 0\end{bmatrix}\delta u+E\delta d\end{aligned}
\]

The voltage input matrix is

\[
{B}_{u}(t)=\begin{bmatrix}{B}_{I}(t) \\ {0}_{2\times 2} \\ {0}_{2\times 2}\end{bmatrix}
\]

where

\[
{B}_{I}(t)={\bar{M}}_{\lambda}^{-1}(t)
\]

The disturbance matrix for

\[
\delta d=\begin{bmatrix}\delta T_{load} \\ \delta F_{load,g}\end{bmatrix}
\]

is

\[
E=\begin{bmatrix}0 & 0 \\ 0 & 0 \\ 0 & 0 \\ -\frac{1}{J} & 0 \\ 0 & 0 \\ 0 & \frac{1}{m_g}\end{bmatrix}
\]

The production measurement perturbations are inherited from the measurement channels defined in Section~\ref{sec:governing-equations}:

\[
\delta\mathscr{y}_I=C_I\delta x\qquad \delta\mathscr{y}_R=C_R\delta x
\]

Although $\delta g$ and $\delta v_g$ are included in the state vector, they are not directly measured in the production signal set.

The key blocks for the subsequent analysis are:

\[
{A}_{Ig}:\delta{x}_{g}\to \delta{\dot{x}}_{I}
\]

\[
{A}_{Rg}:\delta{x}_{g}\to \delta{\dot{x}}_{R}
\]

\[
{A}_{gI}:\delta{x}_{I}\to \delta{\dot{x}}_{g}
\]

The next section uses this linearized model to define the state transition matrix and the finite-time Gramians for current observability, resolver observability, and voltage reachability. The explicit physical interpretation of the coupling blocks is deferred to the local coupling and degeneracy analysis.

\section{State Transition Matrix and Finite-Time Gramians}
\label{sec:state-transition-gramians}

This section uses the linearized perturbation model from Section~\ref{sec:linearization-loaded-trajectory} to define finite-time observability and reachability measures for the air-gap deformation state. The purpose is to evaluate whether the unmeasured air-gap state can be inferred from production current or resolver measurements, and whether it can be influenced by plant-level dq voltage perturbations.

The Gramians in this section are defined for the passive, carrier-off linearized plant. They measure finite-time channel structure between the selected air-gap deformation state, production-accessible measurements, and dq voltage perturbations. They should not be interpreted as closed-loop controller performance metrics or as carrier-on Gramians. Once an injected carrier is introduced, the relevant matrices become carrier-dependent and are defined separately using the carrier-on transition matrix in Section~\ref{sec:carrier-induced-projection}.

For the observability and reachability definitions in this section, disturbance inputs are set to zero. The Gramians defined below should therefore be interpreted as finite-time channel measures of the linearized plant model, not as complete closed-loop performance guarantees.

\subsection{State Transition Matrix}
\label{subsec:state-transition-matrix}

The state transition matrix associated with the homogeneous linearized system

\[
\delta\dot{x}(t)=A(t)\delta x(t)
\]

is defined by

\[
\frac{d}{dt}\Phi(t,t_0)=A(t)\Phi(t,t_0)\qquad \Phi(t_0,t_0)=I_6
\]

This definition does not require $A(t)$ to be constant. Under a frozen-time approximation, $A(t)$ may be replaced locally by $\bar{A}$, but the finite-time formulation itself remains compatible with LTV dynamics over the selected observation window. In this section, however, the matrices correspond to the passive carrier-off plant defined in Section~\ref{sec:linearization-loaded-trajectory}.

For any initial perturbation $\delta x({t}_{0})$, the zero-input state response is

\[
\delta x(t)=\Phi(t,t_0)\delta x(t_0)
\]

Using the state partition

\[
\delta x=\begin{bmatrix}\delta x_I \\ \delta x_R \\ \delta x_g\end{bmatrix}
\]

where

\[
\delta x_I=\begin{bmatrix}\delta I_d \\ \delta I_q\end{bmatrix}\qquad \delta x_R=\begin{bmatrix}\delta\theta_r \\ \delta\omega_r\end{bmatrix}\qquad \delta x_g=\begin{bmatrix}\delta g \\ \delta v_g\end{bmatrix}
\]

the state transition matrix is partitioned as

\[
\Phi(t,t_0)=\begin{bmatrix}\Phi_{II}(t,t_0) & \Phi_{IR}(t,t_0) & \Phi_{Ig}(t,t_0) \\ \Phi_{RI}(t,t_0) & \Phi_{RR}(t,t_0) & \Phi_{Rg}(t,t_0) \\ \Phi_{gI}(t,t_0) & \Phi_{gR}(t,t_0) & \Phi_{gg}(t,t_0)\end{bmatrix}
\]

The state-transition formulation and finite-time input-output maps follow standard linear-system theory \cite{rugh1996linear,kailath1980linear}. The block ${\Phi}_{Ig}\left(t,{t}_{0}\right)$ maps an initial air-gap deformation perturbation into the current state response,

\[
\delta {x}_{I}\left(t\right)={\Phi}_{Ig}\left(t,{t}_{0}\right)\delta {x}_{g}({t}_{0})
\]

when the other initial perturbations are set to zero. Similarly, ${\Phi}_{Rg}\left(t,{t}_{0}\right)$ maps an initial air-gap deformation perturbation into the resolver-state response,

\[
\delta {x}_{R}\left(t\right)={\Phi}_{Rg}\left(t,{t}_{0}\right)\delta {x}_{g}({t}_{0})
\]

For reachability analysis, the block ${\Phi}_{gI}\left({t}_{f},\tau \right)$ maps current-state forcing at time $\tau$ into the final air-gap deformation state at ${t}_{f}$.

\subsection{Current-Based Observability Gramian}
\label{subsec:current-observability-gramian}

The current measurement channel is

\[
\delta\mathscr{y}_I(t)=C_I\delta x(t)=\delta x_I(t)
\]

To evaluate whether the air-gap deformation state $\delta {x}_{g}\left({t}_{0}\right)$ is observable from current residuals over the finite interval $\left[{t}_{0},{t}_{f}\right]$, the unweighted current-based observability Gramian is first defined as

\[
W_{o,g}^{I}(t_0,t_f)=\int_{t_0}^{t_f}\Phi_{Ig}^{T}(t,t_0)\Phi_{Ig}(t,t_0)\,dt
\]

Finite-time observability and reachability Gramians provide standard measures of state-to-output and input-to-state channel conditioning in linear systems \cite{rugh1996linear,kailath1980linear}. This unweighted Gramian characterizes the structural finite-time coupling from the air-gap deformation state to the ${I}_{d}/{I}_{q}$ current response under the linearized model. It does not require any sensor-noise assumption.

For practical feasibility assessment, a calibrated or noise-weighted version can be introduced as

\[
W_{o,g}^{I,R}(t_0,t_f)=\int_{t_0}^{t_f}\Phi_{Ig}^{T}(t,t_0)R_I^{-1}\Phi_{Ig}(t,t_0)\,dt
\]

Here,

\[
R_I=\begin{bmatrix}\sigma_{I_d}^{2} & 0 \\ 0 & \sigma_{I_q}^{2}\end{bmatrix}
\]

is an effective current-residual uncertainty matrix. It is intended as an experimental data interface that can later be populated from current sensor specifications, controller logs, steady-state current residuals, band-limited noise measurements, dq transformation effects, filtering, and inverter nonidealities.

The use of ${R}_{I}$ does not assume that all production current errors are white, stationary, or independent. It only provides a first calibration layer for comparing the air-gap-induced current response against the effective uncertainty level of the residual channel. More detailed colored-noise, filtering, or controller-dependent residual models can be incorporated by replacing this weighting with an experimentally identified residual covariance or frequency-domain weighting.

Thus, ${W}_{o,g}^{I}$ defines the theoretical current-observation structure, while ${W}_{o,g}^{I,R}$ provides a calibration-ready extension for production current-residual uncertainty.

\subsection{Resolver-Based Observability Gramian}
\label{subsec:resolver-observability-gramian}

The resolver measurement channel is

\[
\delta\mathscr{y}_R(t)=C_R\delta x(t)=\delta x_R(t)
\]

The unweighted resolver-based finite-time observability Gramian is defined as

\[
W_{o,g}^{\theta_\omega}(t_0,t_f)=\int_{t_0}^{t_f}\Phi_{Rg}^{T}(t,t_0)\Phi_{Rg}(t,t_0)\,dt
\]

This Gramian measures whether the air-gap deformation state produces a finite-time resolver angle or speed response under the linearized plant model. It is defined separately from the current-based Gramian because the resolver channel does not directly sense flux-linkage perturbations. Instead, it observes the air-gap deformation state only through its effect on electromagnetic torque and rotor mechanical motion.

A calibrated resolver-weighted version is

\[
W_{o,g}^{\theta_\omega,R}(t_0,t_f)=\int_{t_0}^{t_f}\Phi_{Rg}^{T}(t,t_0)R_{\theta_\omega}^{-1}\Phi_{Rg}(t,t_0)\,dt
\]

The resolver uncertainty matrix is

\[
R_{\theta_\omega}=\begin{bmatrix}\sigma_{\theta}^{2} & 0 \\ 0 & \sigma_{\omega}^{2}\end{bmatrix}
\]

This matrix is also an experimental data interface. It can be estimated from resolver resolution, demodulation error, angle quantization, speed-estimator noise, constant-speed residuals, controller logs, or band-limited noise measurements near the frequency of interest. Since ${\omega}_{r}$ is typically estimated from ${\theta}_{r}$, the effective uncertainty in ${\omega}_{r}$ depends strongly on the speed-estimation algorithm and filtering bandwidth.

Therefore, ${W}_{o,g}^{{\theta}_{\omega }}$ represents the ideal finite-time resolver response structure, while ${W}_{o,g}^{{\theta}_{\omega },R}$ represents the same structure after calibration by production resolver uncertainty.

\subsection{Voltage-to-Air-Gap Reachability Gramian}
\label{subsec:voltage-air-gap-reachability-gramian}

The plant-level voltage input enters the perturbation model through

\[
B_u(t)=\begin{bmatrix}B_I(t) \\ 0 \\ 0\end{bmatrix}\qquad B_I(t)=\bar{M}_{\lambda}^{-1}(t)
\]

For zero initial state, the response at final time ${t}_{f}$ due to an input perturbation $\delta u(\tau)$ is

\[
\delta x(t_f)=\int_{t_0}^{t_f}\Phi(t_f,\tau)B_u(\tau)\delta u(\tau)\,d\tau
\]

Selecting only the air-gap deformation state gives

\[
\delta x_g(t_f)=\int_{t_0}^{t_f}\Phi_{gI}(t_f,\tau)B_I(\tau)\delta u(\tau)\,d\tau
\]

The unweighted plant-level voltage-to-air-gap reachability Gramian is defined as

\[
\begin{aligned}W_{r,g}^{V}(t_0,t_f)={}&\int_{t_0}^{t_f}\Phi_{gI}(t_f,\tau)B_I(\tau)B_I^{T}(\tau) \\&\qquad{}\Phi_{gI}^{T}(t_f,\tau)\,d\tau\end{aligned}
\]

This Gramian characterizes the intrinsic plant-level voltage-to-air-gap reachability of the air-gap deformation state. It answers whether dq voltage perturbations have a physical pathway to influence $\delta g$ and $\delta {v}_{g}$  through the motor electrical dynamics and electromagnetic generalized force.

A weighted version is

\[
\begin{aligned}W_{r,g}^{V,R}(t_0,t_f)={}&\int_{t_0}^{t_f}\Phi_{gI}(t_f,\tau)B_I(\tau)R_u^{-1} \\&\qquad{}B_I^{T}(\tau)\Phi_{gI}^{T}(t_f,\tau)\,d\tau\end{aligned}
\]

Here,

\[
R_u=\begin{bmatrix}r_{V_d} & 0 \\ 0 & r_{V_q}\end{bmatrix}
\]

is an input weighting matrix, not a sensor-noise covariance. It represents a calibration-ready interface for actuation cost, voltage authority, available voltage headroom, command uncertainty, or relative weighting between $d$- and $q$-axis voltage perturbations. In a later implementation, ${R}_{u}$ may be informed by DC-bus voltage limits, field-weakening margin, inverter dead time, voltage saturation, current-loop rejection, or command filtering.

Therefore, ${W}_{r,g}^{V}$ defines the theoretical plant-level reachability structure, while ${W}_{r,g}^{V,R}$ provides a practical extension for weighting the usable voltage authority.

This reachability measure remains plant-level. It answers whether a physical voltage-to-current-to-force pathway exists in the linearized plant, not whether a production inverter can supply enough voltage, current, bandwidth, or energy to suppress the selected NVH component. Practical actuation capability is limited by DC-bus voltage, voltage saturation, PWM resolution, current-regulation bandwidth, current-loop rejection, field-weakening margin, thermal and loss budgets, and the possibility that injected perturbations introduce additional torque ripple or electromagnetic noise. Therefore, ${W}_{r,g}^{V}$ should be interpreted as a channel-existence and conditioning measure, not as a guarantee of executable closed-loop attenuation.

\subsection{Summary}
\label{subsec:gramian-summary}

This section defined the state transition matrix and three finite-time Gramians for the air-gap deformation state: the current-based observability Gramian, the resolver-based observability Gramian, and the plant-level voltage-to-air-gap reachability Gramian. The unweighted forms describe the finite-time coupling structure of the linearized plant, while the weighted forms retain interfaces for future experimental calibration through current-residual uncertainty, resolver-estimator uncertainty, and voltage-authority weighting.

These Gramians provide the mathematical objects used in the subsequent local feasibility analysis. Section~\ref{sec:passive-coupling-degeneracy} derives the physical coupling mechanisms behind the Gramian kernels, while Section~\ref{sec:carrier-induced-projection} extends the passive projection framework to carrier-induced observability and reachability.

\section{Local Passive Coupling Conditions and Degeneracy Mechanisms}
\label{sec:passive-coupling-degeneracy}

Section~\ref{sec:state-transition-gramians} defined finite-time observability and reachability Gramians in terms of the state-transition subblocks ${\Phi}_{Ig}$, ${\Phi}_{Rg}$, and ${\Phi}_{gI}$. These Gramians provide a system-level measure of whether the air-gap deformation state can be observed from production current residuals, observed from resolver signals, or influenced by plant-level dq voltage perturbations.

This section derives the local physical coupling conditions that appear inside those Gramian kernels. The analysis is not intended to replace full state-transition-matrix integration. Instead, it identifies the leading-order passive coupling paths that must exist before the corresponding finite-time Gramians can become well-conditioned.

The analysis is performed around a frozen-time nominal loaded operating condition. Let

\[
H={t}_{f}-{t}_{0}
\]

denote a short local analysis horizon. Over this horizon, the linearized dynamics are approximated by

\[
A(t)\approx \bar{A}
\]

The local results retain the direct leading-order coupling paths between current states, resolver states, and air-gap deformation states. Long-horizon recirculation through the full state matrix, damping and resonance effects, controller dynamics, inverter nonidealities, disturbance forcing, measurement noise, and input constraints are not included in the local derivation. These effects must be evaluated through the complete finite-time framework or future numerical validation.

The resulting conditions should therefore be interpreted as passive feasibility boundaries. A nonzero local coupling does not guarantee practical observability or usable control authority, while a locally degenerate passive path indicates that additional mechanisms, such as calibrated nonideal asymmetry or carrier-induced projection, may be required.

\subsection{From Finite-Time Gramians to Local Coupling Blocks}
\label{subsec:gramians-to-local-coupling}

The current-based finite-time Gramian is

\[
{W}_{o,g}^{I}\left({t}_{0},{t}_{f}\right)={\int}_{{t}_{0}}^{{t}_{f}}{\Phi}_{Ig}^{T}\left(t,{t}_{0}\right){\Phi}_{Ig}\left(t,{t}_{0}\right)dt
\]

The subblock ${\Phi}_{Ig}$ maps an initial air-gap deformation perturbation into the current-state response. For a short horizon, with $t={t}_{0}+\tau$,

\[
{\Phi}_{Ig}\left({t}_{0}+\tau,{t}_{0}\right)={A}_{Ig}\tau +O({\tau}^{2})
\]

where

\[
{A}_{Ig}={\left.\frac{\partial {\dot{x}}_{I}}{\partial {x}_{g}}\right\vert}_{\bar{x},\bar{u}}
\]

Therefore,

\[
W_{o,g}^{I}=\frac{H^3}{3}A_{Ig}^{T}A_{Ig}+O(H^4)
\]

Similarly, the weighted current Gramian has the local form

\[
W_{o,g}^{I,R}=\frac{H^3}{3}A_{Ig}^{T}R_I^{-1}A_{Ig}+O(H^4)
\]

Thus, local current observability is controlled by the coupling block ${A}_{Ig}$.

For resolver-based observation, the relevant local coupling block is

\[
{A}_{Rg}={\left.\frac{\partial {\dot{x}}_{R}}{\partial {x}_{g}}\right\vert}_{\bar{x},\bar{u}}
\]

Unlike the current channel, resolver response arises through torque perturbation and rotor integration, so ${\Phi}_{Rg}$ includes additional time-integration structure. Nevertheless, the local resolver Gramian is ultimately controlled by whether ${A}_{Rg}$ contains a nonzero torque-projection path from the air-gap deformation state.

For voltage-to-air-gap reachability, the local reachability kernel is governed by the cascade

\[
\delta u\to \delta {x}_{I}\to \delta {x}_{g}
\]

where

\[
{B}_{I}={\left.\frac{\partial {\dot{x}}_{I}}{\partial u}\right\vert}_{\bar{x},\bar{u}}\qquad {A}_{gI}={\left.\frac{\partial {\dot{x}}_{g}}{\partial {x}_{I}}\right\vert}_{\bar{x},\bar{u}}
\]

Consequently, the finite-time Gramians reduce locally to three physical coupling questions:

\[
{A}_{Ig}\neq 0\qquad {A}_{Rg}\neq 0\qquad {A}_{gI}{B}_{I}\neq 0
\]

The remainder of this section derives these blocks from the co-energy model.

\subsection{Current-Based Observation Path}
\label{subsec:current-observation-path}

The current-based observation path is governed by

\[
{A}_{Ig}={\left.\frac{\partial {\dot{x}}_{I}}{\partial {x}_{g}}\right\vert}_{\bar{x},\bar{u}}
\]

which maps air-gap deformation perturbations into current perturbation dynamics,

\[
\delta {x}_{g}\to \delta {\dot{x}}_{I}
\]

From the linearized current dynamics,

\[
\delta\dot{x}_I=\bar{M}_{\lambda}^{-1}\delta r
\]

the leading-order dependence of the residual vector on $\delta g$ and $\delta {v}_{g}$ gives

\[
{A}_{Ig}={\bar{M}}_{\lambda }^{-1}\begin{bmatrix}p\Omega {\bar{\lambda }}_{q,g} & -{\bar{\lambda }}_{d,g} \\ -p\Omega {\bar{\lambda }}_{d,g} & -{\bar{\lambda }}_{q,g}\end{bmatrix}
\]

Let

\[
{\bar{\lambda }}_{d,g}=a\qquad {\bar{\lambda }}_{q,g}=b
\]

The determinant of inner coupling matrix is

\[
\det\begin{bmatrix}p\Omega b & -a \\ -p\Omega a & -b\end{bmatrix}=-p\Omega(a^2+b^2)
\]

Since

\[
\det\!\left(\bar{M}_{\lambda}\right)\neq 0
\]

the block ${A}_{Ig}$ has full column rank when

\[
\Omega \neq 0\qquad {\ell}_{\lambda }\neq 0
\]

Thus, a nonzero flux-linkage sensitivity vector provides a first-order current observation path from the air-gap deformation state to the ${I}_{d}/{I}_{q}$ current residuals.

For a short horizon,

\[
W_{o,g}^{I}\approx\frac{H^3}{3}A_{Ig}^{T}A_{Ig}
\]

If

\[
{\ell}_{\lambda }=0
\]

then

\[
{A}_{Ig}=0
\]

and the leading-order current observation path collapses. In that case, the air-gap deformation may still exist mechanically, but it does not project into dq current residuals at first order under the passive electromagnetic closure.

\subsection{Resolver Torque-Projection Path}
\label{subsec:resolver-torque-projection-path}

The resolver channel observes the air-gap deformation state through a different physical mechanism. It does not measure local flux redistribution directly. Instead, the deformation must perturb electromagnetic torque, and that torque perturbation must propagate into rotor speed and rotor angle.

The relevant block is

\[
{A}_{Rg}={\left.\frac{\partial {\dot{x}}_{R}}{\partial {x}_{g}}\right\vert}_{\bar{x},\bar{u}}
\]

which maps air-gap deformation perturbations into resolver-state dynamics,

\[
\delta {x}_{g}\to \delta {\dot{x}}_{R}
\]

From the rotor mechanical dynamics, the direct air-gap contribution is governed by ${\bar{T}}_{e,g}$. Using the co-energy-based torque expression,

\[
{T}_{e}=\frac{3}{2}\left[p\left({\lambda}_{d}{I}_{q}-{\lambda}_{q}{I}_{d}\right)+{\mathscr{w}}_{c,{\theta}_{r}}\right]
\]

\[
{\bar{T}}_{e,g}=\frac{3}{2}\left[p\left({\bar{\lambda }}_{d,g}{I}_{q0}-{\bar{\lambda }}_{q,g}{I}_{d0}\right)+{\bar{\mathscr{w}}}_{c,{\theta}_{r},g}\right]
\]

Define

\[
{a}_{T}=\frac{{\bar{T}}_{e,g}}{J}=\frac{3}{2J}\left[p\left({\bar{\lambda }}_{d,g}{I}_{q0}-{\bar{\lambda }}_{q,g}{I}_{d0}\right)+{\bar{\mathscr{w}}}_{c,{\theta}_{r},g}\right]
\]

The leading-order resolver coupling block is

\[
{A}_{Rg}=\begin{bmatrix}0 & 0 \\ {a}_{T} & 0\end{bmatrix}
\]

The corresponding torque-projection margin is

\[
\rho_T=\left|p\left(\bar{\lambda}_{d,g}I_{q0}-\bar{\lambda}_{q,g}I_{d0}\right)+\bar{\mathscr{w}}_{c,\theta_r,g}\right|
\]

Resolver observability degenerates when

\[
\rho_T\to 0
\]

This can occur even when ${\ell}_{\lambda }\neq 0$. Under the baseline symmetry approximation,

\[
{\bar{\mathscr{w}}}_{c,{\theta}_{r},g}\approx 0
\]

so the torque-projection margin reduces to

\[
\rho_T\approx p\left|\bar{\lambda}_{d,g}I_{q0}-\bar{\lambda}_{q,g}I_{d0}\right|
\]

Thus, resolver visibility depends not only on the existence of flux-linkage sensitivity, but also on whether that sensitivity produces torque perturbation at the selected current operating point.

For short-time analysis,

\[
{\Phi}_{Rg}\left({t}_{0}+\tau,{t}_{0}\right)\approx {a}_{T}\begin{bmatrix}{1}/{2}{\tau}^{2} & {1}/{6}{\tau}^{3} \\ \tau & {1}/{2}{\tau}^{2}\end{bmatrix}
\]

Therefore, the resolver observability Gramian is locally proportional to ${a}_{T}^{2}$. The resolver channel loses leading-order sensitivity as

\[
{a}_{T}\to 0
\]

This result explains why current-based observability and resolver-based observability are not equivalent. Current residuals depend directly on ${\ell}_{\lambda }$, whereas resolver response depends on the torque-projection margin $\rho_T$.

This resolver path should be interpreted as a theoretical torque-to-motion observation channel. In practice, rotor inertia and speed-estimator filtering can strongly attenuate medium- and high-frequency torque perturbations before they appear in ${\omega}_{r}$ or ${\theta}_{r}$. Therefore, resolver-based observability may be poorly conditioned at high speed or high order, even when

\[
\rho_T\neq 0
\]

This limitation motivates treating resolver sensing as a separate and potentially weak observation channel, rather than assuming that torque projection automatically provides a usable production measurement. It also motivates the electrical-residual and carrier-sideband channels developed later.

\subsection{Voltage-to-Air-Gap Reachability Path}
\label{subsec:voltage-air-gap-reachability-path}

The plant-level voltage input first perturbs dq currents through the electrical dynamics. Those current perturbations then generate generalized electromagnetic force on the air-gap deformation coordinate. Therefore, the voltage-to-air-gap reachability path is the cascade

\[
\delta u\to \delta {x}_{I}\to \delta {x}_{g}
\]

From the linearized air-gap deformation dynamics,

\[
\delta\dot{v}_g=\frac{1}{m_g}\left(\delta F_{e,g}-c_g\delta v_g-k_g\delta g+\delta F_{load,g}\right)
\]

and

\[
{F}_{e,g}={s}_{g}\frac{3}{2}{\mathscr{w}}_{c,g}
\]

the current derivatives of generalized electromagnetic force are

\[
{F}_{e,g,{I}_{d}}={s}_{g}\frac{3}{2}{\bar{\lambda }}_{d,g}\qquad {F}_{e,g,{I}_{q}}={s}_{g}\frac{3}{2}{\bar{\lambda }}_{q,g}
\]

Therefore,

\[
{A}_{gI}=\begin{bmatrix}0 & 0 \\ \gamma {\bar{\lambda }}_{d,g} & \gamma {\bar{\lambda }}_{q,g}\end{bmatrix}\qquad \gamma =\frac{3{s}_{g}}{2{m}_{g}}
\]

This expression shows that the same flux-linkage sensitivity vector ${\ell}_{\lambda }$ that governs current-based observability also governs the passive voltage-to-air-gap reachability path.

For a local reachability approximation, let

\[
\Delta ={t}_{f}-\tau
\]

The leading-order map from current perturbation at time $\tau$ to final air-gap deformation state is

\[
{\Phi}_{gI}\left({t}_{f},\tau \right)\approx \gamma \begin{bmatrix}\frac{1}{2}{\Delta}^{2} \\ \Delta \end{bmatrix}{\ell}_{\lambda }^{T}
\]

Combining this with the voltage-to-current map gives

\[
{\Phi}_{gI}\left({t}_{f},\tau \right){B}_{I}\approx \gamma \begin{bmatrix}\frac{1}{2}{\Delta}^{2} \\ \Delta \end{bmatrix}{\ell}_{\lambda }^{T}{\bar{M}}_{\lambda }^{-1}
\]

The unweighted local voltage reachability Gramian becomes

\[
\begin{aligned}W_{r,g}^{V}\approx{}&\gamma^2\ell_{\lambda}^{T}\bar{M}_{\lambda}^{-1}\bar{M}_{\lambda}^{-T}\ell_{\lambda} \\&\quad\begin{bmatrix}H^5/20 & H^4/8 \\ H^4/8 & H^3/3\end{bmatrix}\end{aligned}
\]

Since ${M}_{\lambda }$ is invertible, the unweighted scalar is positive whenever

\[
{\ell}_{\lambda }\neq 0
\]

Therefore, under the plant-level passive model, a nonzero flux-linkage sensitivity vector provides a first-order voltage-to-air-gap reachability path.

This result should be interpreted as a local channel-existence condition. It does not imply that the production inverter has sufficient voltage, current, bandwidth, or thermal margin to generate a practically useful air-gap force. Those implementation limits are outside the passive local derivation and must be evaluated through input weighting, controller modeling, and experimental validation.

The three local paths derived above show that passive current observation and passive voltage reachability share a common bottleneck: the flux-linkage sensitivity vector ${\ell}_{\lambda }$. Resolver visibility, in contrast, depends on the torque-projection margin $\rho_T$.

\subsection{Physical Interpretation: Global Projection and Harmonic Orthogonality}
\label{subsec:global-projection-harmonic-orthogonality}

The preceding local conditions show that ${\ell}_{\lambda }$ is the common bottleneck for passive current observation and voltage reachability. The remaining question is whether this vector is expected to be nonzero for the selected low-order rotor-coupled air-gap deformation. This requires examining the electromagnetic projection that generates ${\ell}_{\lambda }$.

A key limitation of production-signal-based observability is that ${I}_{d}/{I}_{q}$ does not measure local air-gap flux density. The controller observes winding-level electrical quantities after spatial integration through the three-phase winding distribution and dq transformation. Therefore, a local air-gap flux redistribution may be mechanically significant while producing little or no net first-order change in dq current.

This distinction can be expressed as a projection problem. A local air-gap deformation modifies the permeance field,

\[
\delta {\Lambda}_{g}(\beta,t)
\]

but the current channel can only sense its projection through a dq electromagnetic kernel,

\[
\delta\lambda_i(t)\sim\int_{0}^{2\pi}K_i(\beta)\delta\Lambda_g(\beta,t)\,d\beta\qquad i\in\{d,q\}
\]

Thus, local flux-density variation is not equal to observable dq flux-linkage variation. Only the component of the air-gap deformation that survives this global projection can appear in ${I}_{d}/{I}_{q}$ residuals.

For a selected low-order deformation mode, write the normalized spatial deformation shape as

\[
{\phi}_{r}\left(\beta \right)=\cos(r\beta +{\phi}_{g})
\]

with

\[
r\in \{1,2,3,4\}
\]

the flux-linkage sensitivity vector is generated by the physical chain

\[
g\to {\Lambda}_{g}\to {\psi}_{i},{L}_{i,j}\to {\lambda}_{d}{,\lambda}_{q}
\]

At the nominal loaded operating condition,

\[
{\bar{\lambda }}_{d,g}={\bar{\psi }}_{d,g}+{\bar{L}}_{dd,g}{I}_{d0}+{\bar{L}}_{dq,g}{I}_{q0}
\]

\[
{\bar{\lambda }}_{q,g}={\bar{\psi }}_{q,g}+{\bar{L}}_{dq,g}{I}_{d0}+{\bar{L}}_{qq,g}{I}_{q0}
\]

Using

\[
{\left.\frac{\partial {\Lambda}_{g}}{\partial g}\right\vert}_{g=0}=-\frac{{\mu}_{0}}{{g}_{0}^{2}}{\phi}_{r}(\beta)
\]

the PM flux-linkage sensitivities and the inductance sensitivities are

\[
{\bar{\psi }}_{i,g}=-\frac{{K}_{\psi }{\mu}_{0}}{{g}_{0}^{2}}{\int}_{0}^{2\pi }{b}_{i}(\beta){\mathscr{F}}_{m}(\beta){\phi}_{r}(\beta)d\beta
\]

\[
i\in \left\{d,q\right\}
\]

\[
{\bar{L}}_{ij,g}=-\frac{{K}_{L}{\mu}_{0}}{{g}_{0}^{2}}{\int}_{0}^{2\pi }{b}_{i}(\beta){b}_{j}(\beta){\phi}_{r}(\beta)d\beta
\]

\[
i,j\in \left\{d,q\right\}
\]

Therefore,

\[
\begin{aligned}\bar{\lambda}_{d,g}=-\frac{\mu_0}{g_0^2}\Bigl[{}&K_{\psi}\int_{0}^{2\pi}b_d(\beta)\mathscr{F}_m(\beta)\phi_r(\beta)\,d\beta \\&+K_L I_{d0}\int_{0}^{2\pi}b_d^2(\beta)\phi_r(\beta)\,d\beta \\&+K_L I_{q0}\int_{0}^{2\pi}b_d(\beta)b_q(\beta)\phi_r(\beta)\,d\beta\Bigr]\end{aligned}
\]

\[
\begin{aligned}\bar{\lambda}_{q,g}=-\frac{\mu_0}{g_0^2}\Bigl[{}&K_{\psi}\int_{0}^{2\pi}b_q(\beta)\mathscr{F}_m(\beta)\phi_r(\beta)\,d\beta \\&+K_L I_{d0}\int_{0}^{2\pi}b_d(\beta)b_q(\beta)\phi_r(\beta)\,d\beta \\&+K_L I_{q0}\int_{0}^{2\pi}b_q^2(\beta)\phi_r(\beta)\,d\beta\Bigr]\end{aligned}
\]

These equations show that ${\ell}_{\lambda }$ is determined by the spatial overlap among

\[
{g}_{0}\qquad {\phi}_{r}\left(\beta \right)\qquad {b}_{d}\left(\beta \right),{b}_{q}\left(\beta \right)\qquad {\mathscr{F}}_{m}(\beta)\qquad {I}_{d0},{I}_{q0}
\]

The PM terms determine whether the selected low-order air-gap deformation projects into PM-induced dq flux linkage. The inductance terms determine whether the same deformation changes the current-produced magnetic field. The loaded currents ${I}_{d0},{I}_{q0}$ enter through the current-loaded inductance contribution.

Under an ideal symmetric electromagnetic closure, the integrals defining ${\ell}_{\lambda }$ are full-circumference projections. A nonzero integral requires that the product being integrated contain a zero-order spatial component. Otherwise, the positive and negative spatial contributions cancel over one mechanical revolution.

For the PM flux-linkage sensitivity,

\[
{\bar{\psi }}_{i,g}\propto {\int}_{0}^{2\pi }{b}_{i}(\beta){\mathscr{F}}_{m}(\beta){\phi}_{r}(\beta)d\beta
\]

Using the baseline dq basis,

\[
{b}_{d}\left(\beta \right)=\cos(p\beta)
\]

\[
{b}_{q}\left(\beta \right)=\sin(p\beta)
\]

and the PM MMF harmonic expansion,

\[
{\mathscr{F}}_{m}\left(\beta \right)={\sum}_{n}{F}_{mn}\cos(np\beta +{\varphi}_{mn})
\]

while the selected low-order deformation is

\[
{\phi}_{r}\left(\beta \right)=\cos(r\beta +{\phi}_{g})
\]

the PM flux-linkage projection contains spatial orders of the form

\[
\pm p\pm np\pm r
\]

A nonzero full-circumference projection requires one of these combinations to produce a zero-order term,

\[
\pm p\pm np\pm r=0
\]

Equivalently,

\[
p(\pm 1\pm n)\pm r=0
\]

This interpretation is consistent with the use of spatial harmonic projection and force-harmonic analysis in PMSM electromagnetic modeling and NVH studies \cite{soresini2024noise,wang2023modelling,krause2013analysis,krishnan2010permanent,yang2020radial}. This condition provides a general harmonic-matching criterion for PMSM architectures. It shows that passive dq visibility is not determined merely by the existence of a mechanical deformation mode, but by whether that mode generates a zero-order projection after multiplication by the winding and PM field harmonics.

For a broad class of multi-pole PMSMs, low-order mechanical deformation modes satisfy

\[
0<r<p
\]

In that case, the term $p(\pm 1\pm n)$ is an integer multiple of $p$, and a low-order $r<p$ term cannot generally cancel it. Therefore, under the ideal symmetric harmonic closure, a low-order rotor-coupled air-gap deformation with
$0<r<p$  is not expected to produce a first-order PM flux-linkage projection.

For the 48-slot, 8-pole numerical example considered in Section~\ref{sec:numerical-feasibility}, $p=4$, and thus $r=4$ constitutes a boundary case rather than satisfying the strict condition $0<r<p$. It is nevertheless retained in the numerical sweep because the study evaluates the prescribed low-order set $r\in\{1,2,3,4\}$ and because $r=4$ remains within the selected low-order deformation subspace. Under the specific PM-harmonic and inductance-projection closure adopted in Section~\ref{sec:numerical-feasibility}, the corresponding projection also remains below the numerical floor.

The same argument applies to the inductance sensitivity terms. Products such as

\[
{b}_{d}^{2}\left(\beta \right)\qquad {b}_{q}^{2}\left(\beta \right)\qquad {b}_{d}\left(\beta \right){b}_{q}\left(\beta \right)
\]

generate spatial components at order 0 and 2p. Multiplication by the selected deformation mode ${\phi}_{r}\left(\beta \right)$ produces spatial orders

\[
0\pm r\qquad 2p\pm r
\]

A zero-order component would require either

\[
r=0\qquad r=2p
\]

For nonzero low-order deformation modes satisfying

\[
0<r<2p
\]

these terms do not contain a zero-order component. Therefore, the ideal first-order inductance sensitivity to a low-order air-gap deformation also vanishes under the same symmetric closure.

Consequently, for a broad class of multi-pole PMSMs under the ideal symmetric closure,

\[
{\bar{\psi }}_{i,g}\approx 0\qquad {\bar{L}}_{ij,g}\approx 0
\]

and therefore

\[
{\ell}_{\lambda }\approx 0
\]

This result is a passive orthogonality boundary, not a universal statement that real machines have exactly zero coupling. Slotting, saturation, winding-distribution effects, skew, finite stack length, end effects, rotor or stator asymmetry, dynamic eccentricity, and manufacturing variation may create small but nonzero effective projection coefficients. Such effects are typically captured through detailed electromagnetic or coupled electromagnetic-structural simulation rather than ideal full-circumference harmonic closure \cite{soresini2024noise,wang2023modelling,yang2020radial,muller2025evaluation}. In that case, the finite-time Gramians from Section~\ref{sec:state-transition-gramians} can quantify the resulting weak passive observability or reachability once the effective maps are calibrated.

The framework applies to arbitrary selected deformation order $r$, but passive degeneracy occurs whenever the spatial harmonic content of the mechanical deformation does not generate a zero-order component after projection through the dq electromagnetic kernel.

A rotor-coupled low-order deformation can therefore be mechanically significant while remaining first-order invisible to production dq electrical signals if its spatial deformation pattern is orthogonal to the dq electromagnetic sensitivity. The mode is mechanically present, but electrically invisible at first order under the passive symmetric projection.

\subsection{Feasibility Boundary and Scope}
\label{subsec:feasibility-boundary-scope}

The central implication of the local analysis is that the existence of a low-order rotor-coupled air-gap deformation mode does not by itself imply production-signal observability. For a selected deformation order $r$, first-order passive current observability and plant-level voltage reachability require a nonzero electromagnetic projection,

\[
\ell_{\lambda}=\begin{bmatrix}\bar{\lambda}_{d,g} \\ \bar{\lambda}_{q,g}\end{bmatrix}\neq 0
\]

If the selected mode is orthogonal to the dq electromagnetic projection, then

\[
{\ell}_{\lambda }\approx 0
\]

and the leading-order current-observation and voltage-reachability paths collapse:

\[
{A}_{Ig}\approx 0\qquad {A}_{gI}\approx 0
\]

In that case, the deformation mode may be mechanically significant while remaining first-order invisible to passive production current signals. Similarly, resolver visibility requires a nonzero torque projection,

\[
\rho_T=\left|p\left(\bar{\lambda}_{d,g}I_{q0}-\bar{\lambda}_{q,g}I_{d0}\right)+\bar{\mathscr{w}}_{c,\theta_r,g}\right|
\]

and degenerates as

\[
\rho_T\to 0
\]

Thus, current visibility, resolver visibility, and voltage reachability are related but not equivalent. Current observation and voltage reachability share the same electromagnetic projection bottleneck, ${\ell}_{\lambda }$, whereas resolver observation depends on the torque-projection margin $\rho_T$.

This result should not be interpreted as a weakness of the modeling framework. Rather, it identifies the condition under which passive production-signal-based observation and reachability are physically unavailable at first order. If calibrated magnetic maps or experimental residuals produce a nonzero effective ${\ell}_{\lambda }$, the Gramians from Section~\ref{sec:state-transition-gramians} quantify the resulting observability and reachability. If ${\ell}_{\lambda }$ remains near zero, the framework predicts that passive production electrical signals are insufficient for the selected mode.

The conditions derived in this section are local, leading-order passive feasibility conditions. They identify the dominant physical couplings that enter the finite-time Gramians, but they do not replace full numerical integration of the state transition matrix. Long-horizon modal recirculation, damping and resonance effects, controller dynamics, inverter nonidealities, disturbance forcing, magnetic-map uncertainty, measurement noise, and input constraints must be evaluated through the complete finite-time framework or future validation.

This passive feasibility boundary motivates the next step. If a selected low-order rotor-coupled deformation mode is first-order invisible under the passive dq projection, then passive sensing alone may be insufficient. The next section therefore examines whether active symmetry-breaking mechanisms, such as saturation-assisted carriers or intentional harmonic injection, can create an additional carrier-dependent measurement or force projection for an otherwise passive-invisible mode.

The transition from Section~\ref{sec:passive-coupling-degeneracy} to Section~\ref{sec:carrier-induced-projection} should therefore be understood as a shift from passive projection analysis to active projection design. Section~\ref{sec:passive-coupling-degeneracy} identifies when the unmodified production signal set may be poorly conditioned. Section~\ref{sec:carrier-induced-projection} asks whether a physically realizable carrier can modify the electromagnetic projection itself, producing new carrier-dependent coefficients ${\Gamma}_{y}\left(m\right)$ or ${\Gamma}_{F}\left(m\right)$.

\section{Carrier-Induced Electromagnetic Projection for Observability and Reachability}
\label{sec:carrier-induced-projection}

This section examines whether an active carrier can modify the projection problem. The central idea is not to directly measure local air-gap flux density. The controller cannot do that with production signals. Instead, the objective is to inject a known electromagnetic carrier that creates a carrier-dependent spatial weighting or force component. If this carrier interacts with the selected deformation mode, the otherwise passive-invisible mode may become visible through measurable sidebands or reachable through a carrier-induced generalized force.

The carrier introduced here should not be interpreted as a conventional sensorless-position injection signal. High-frequency signal injection and carrier-based methods are well established for saliency-based or saturation-assisted sensorless position estimation \cite{corley1998rotor,jebai2012signal,jebai2012sensorless}. Its role is not to estimate rotor position, classify eccentricity faults, or directly cancel a premeasured harmonic response. Instead, the carrier is treated as an active electromagnetic projection mechanism that may create additional measurement or force channels for the selected air-gap deformation state. The key question is whether the injected carrier can change the projection structure from the passive coefficients ${\ell}_{\lambda }$ and $\rho_T$ to carrier-dependent coefficients ${\Gamma}_{y}\left(m\right)$ and ${\Gamma}_{F}\left(m\right)$.

Once the carrier is injected, the relevant system is generally not the same passive linearized plant used in Sections~\ref{sec:linearization-loaded-trajectory}--\ref{sec:passive-coupling-degeneracy}. The carrier-on air-gap subsystem and residual channel are generally LTV, or LTP when the carrier is rotor-synchronous and the operating speed is locally constant. Therefore, the carrier-induced Gramians below are defined using carrier-on quantities such as $A_{g,c}\left(t\right)$, $C_{g,c}\left(t\right)$, $B_{g,c}\left(t\right)$, and ${\Phi}_{g,c}$. The reduced analytical expressions in this section expose the projection mechanism; complete carrier-on Floquet, harmonic-balance, or full finite-time LTV/LTP evaluation is left for future numerical validation.

\subsection{Passive Degeneracy and the Need for Carrier-Induced Projection}
\label{subsec:passive-degeneracy-carrier-projection}

The passive first-order flux-linkage perturbation has the form

\[
\delta\lambda_i(t)\approx\bar{\lambda}_{i,g}g(t)\qquad i\in\{d,q\}
\]

When the passive projection degenerates,

\[
{\bar{\lambda }}_{i,g}\approx 0
\]

the current channel loses first-order sensitivity to the selected deformation mode. This is the passive invisibility result derived in Section~\ref{sec:passive-coupling-degeneracy}.

However, this result only concerns the unmodified dq projection. If an active carrier with amplitude $\epsilon$ is introduced, the flux linkage becomes a function of both the deformation coordinate and the carrier:

\[
{\lambda}_{i}={\lambda}_{i}(g,\epsilon,t)
\]

The use of saturation-aware or energy-based models in signal-injection analysis motivates treating carrier response as a local mixed or bilinear electromagnetic effect \cite{jebai2012signal,jebai2012sensorless,jebai2014energy,jebai2011estimation}. A local expansion gives

\[
\begin{aligned}\lambda_i(g,\epsilon,t)={}&\lambda_i(0,0)+\lambda_{i,\varepsilon}\epsilon c(t)+\lambda_{i,g}g(t) \\&+\lambda_{i,g\varepsilon}g(t)\epsilon c(t)+O(g^2+\epsilon^2)\end{aligned}
\]

If the passive channel is degenerate, the mixed term may still remain:

\[
\delta\lambda_i^{mix}(t)\approx\lambda_{i,g\varepsilon}\epsilon g(t)c(t)
\]

Thus,

\[
\lambda_{i,g}=0\nRightarrow\lambda_{i,g\varepsilon}=0
\]

This is the theoretical basis of carrier-assisted observability. Passive invisibility rules out the unmodified first-order dq projection, but it does not rule out bilinear coupling between the deformation and an injected carrier.

The expansion above is a local small-signal statement. It does not imply that arbitrarily large carrier amplitudes are acceptable or that all carrier-induced sidebands are useful. The carrier amplitude must remain within a regime where the mixed projection is identifiable, while inverter limits, current-loop behavior, thermal loss, and additional electromagnetic noise remain acceptable.

\subsection{Carrier-Induced Low-Order Saturation Masks}
\label{subsec:carrier-saturation-masks}

The controller cannot directly prescribe an arbitrary spatial weighting function around the air gap. It can only command voltages or currents through the available three-phase winding and inverter. Therefore, a useful carrier must be physically realizable by the motor winding and must create a nonzero spatial mixing component through saliency, saturation, slotting, winding harmonics, or other machine nonlinearities.

Let the dominant electromagnetic field have spatial order $p$, corresponding to the pole-pair number. In simplified form, the main field may be written as

\[
{B}_{p}\left(\beta,t\right)={B}_{p}\cos(p\beta -{\omega}_{e}t)
\]

Assume that the controller injects a small carrier that produces a spatial field component of order $m$,

\[
{B}_{c}\left(\beta,t\right)=\epsilon {B}_{m}\cos(m\beta -{\omega}_{c}t+{\phi}_{c})
\]

The total field is approximately

\[
\begin{aligned}B(\beta,t)={}&B_p\cos(p\beta-\omega_e t) \\&+\epsilon B_m\cos(m\beta-\omega_c t+\phi_c)\end{aligned}
\]

Magnetic saturation and Maxwell stress depend nonlinearly on the local field amplitude. A simple way to see the spatial mixing is to consider the cross term in ${B}^{2}$:

\[
B^2\supset 2\epsilon B_pB_m\cos(p\beta-\omega_e t)\cos(m\beta-\omega_c t+\phi_c)
\]

Using the product identity, the cross term contains spatial orders

\[
m+p\qquad \lvert m-p\rvert
\]

Therefore, if the carrier order satisfies

\[
\lvert m-p\rvert=r
\]

or equivalently,

\[
m=p\pm r
\]

then the carrier and the main field generate an $r$-order spatial beating pattern. The temporal frequency of this beating pattern depends on the carrier implementation and the reference frame, so it is represented compactly by a known carrier waveform $c(t)$.

This field-mixing relation is a necessary spatial-matching condition in the simplified analytical picture, not a sufficient guarantee of useful projection. The winding distribution, inverter bandwidth, current-regulation bandwidth, voltage headroom, saturation level, skew, slotting, and magnetic-map nonlinearity determine whether the commanded carrier actually produces a usable $m$-order electromagnetic component. Therefore, the carrier must belong to the physically available carrier set,

\[
m\in {\mathscr{M}}_{act}
\]

where ${\mathscr{M}}_{act}$ depends on winding factors, inverter bandwidth, current regulation bandwidth, voltage headroom, and allowed harmonic injection. Carrier realizability and useful injected-signal response are limited by winding saliency, saturation, inverter dynamics, and current-loop implementation, as also emphasized in signal-injection and harmonic-current studies \cite{corley1998rotor,jebai2012signal,jebai2012sensorless,jahns1996pulsating,yan2019torque,erken2016online,dai2026data,soresini2025numerical}.

This $r$-order beating pattern acts as a carrier-induced saturation mask or force mask:

\[
{W}_{sat}\left(\beta,t\right)=1+\epsilon {W}_{r}\cos(r\beta +{\phi}_{c})c(t)+O({\epsilon}^{2})
\]

This is the physical mechanism by which the controller may create useful asymmetry. The controller does not directly impose a local air-gap sensor. It injects a realizable electromagnetic carrier. The motor’s nonlinear magnetic field converts the interaction between the main $p$-order field and the carrier $m$-order field into a low-order spatial mask.

Here, $m$ denotes the spatial order of the carrier-induced field component. Later, $h$ denotes the order-domain or time-domain carrier order used in residual detection. The mapping between $h$ and $m$ depends on winding distribution, reference frame, inverter dynamics, and current-loop behavior.

\subsection{Carrier-Dependent Observation Coefficient}
\label{subsec:carrier-observation-coefficient}

The carrier-induced saturation mask changes the measurement projection. Under passive sensing, the selected deformation is projected through an unmodified dq kernel,

\[
{\int}_{0}^{2\pi }{K}_{i}(\beta){\phi}_{r}(\beta)d\beta
\]

If this projection vanishes, passive current observability degenerates.

With a carrier-induced mask, the mixed projection becomes

\[
\Gamma_{y,i}(m)=\int_{0}^{2\pi}W_m(\beta)K_i(\beta)\phi_r(\beta)\,d\beta\qquad i\in\{d,q\}
\]

where ${W}_{m}(\beta)$ is the effective spatial weighting produced by the injected carrier of order $m$. The active observation condition is

\[
{\Gamma}_{\mathscr{y},i}\left(m\right)\neq 0
\]

for at least one measured electrical channel $i$.

If this condition holds, the mixed flux-linkage perturbation has the form

\[
\delta {\lambda}_{i}^{mix}\left(t\right)\approx \epsilon {\Gamma}_{\mathscr{y},i}\left(m\right)g(t)c\left(t\right)
\]

Define the carrier-dependent observation vector

\[
{\Gamma}_{\mathscr{y}}\left(m\right)=\begin{bmatrix}{\Gamma}_{\mathscr{y},d}(m) \\ {\Gamma}_{\mathscr{y},q}(m)\end{bmatrix}
\]

Then the carrier-induced mixed flux perturbation can be written compactly as

\[
\delta {\lambda}^{mix}(t)\approx \epsilon {\Gamma}_{\mathscr{y}}\left(m\right)g(t)c(t)
\]

The time-frequency signature of this mixed term is discussed in Section~\ref{subsec:sideband-detection}.

\subsection{Carrier-Dependent Output Equation and Observability Gramian}
\label{subsec:carrier-output-observability-gramian}

To make the carrier-induced observation mechanism precise, a specific production-accessible residual channel is selected. The primary channel in this framework is the carrier-synchronous dq voltage-equation residual, constructed from

\[
V_d^{cmd}\qquad V_q^{cmd}\qquad I_d\qquad I_q\qquad\theta_r\qquad\omega_r
\]

This channel is preferred over raw current ripple because the current regulator may suppress current perturbations while the required compensation remains visible in voltage command, current-loop effort, or model residual form.

Let

\[
{\mathscr{y}}_{c}\left(t\right)=\begin{bmatrix}{r}_{d,c}(t) \\ {r}_{q,c}(t)\end{bmatrix}
\]

denote the carrier-synchronous dq residual after subtracting the nominal voltage-equation prediction. The known carrier-only response is also removed, giving

\[
{\tilde{\mathscr{y}}}_{c}\left(t\right)={\mathscr{y}}_{c}\left(t\right)-{\hat{\mathscr{y}}}_{0}\left(t\right)-{\hat{\mathscr{y}}}_{\epsilon }\left(t\right)
\]

where ${\hat{\mathscr{y}}}_{0}\left(t\right)$ is the nominal residual and ${\hat{\mathscr{y}}}_{\epsilon }\left(t\right)$ is the residual produced by the injected carrier in the absence of the target deformation.

With the carrier on, the local air-gap subsystem is generally LTV or LTP:

\[
\dot{x}_g(t)=A_{g,c}(t)x_g(t)+\cdots
\]

Periodic-system and LTV/LTP formulations provide the appropriate framework for analyzing carrier-on dynamics when the injected signal is rotor-synchronous or periodically modulated \cite{rugh1996linear,kailath1980linear,bittanti2009periodic}. The carrier-on state-transition matrix satisfies

\[
\dot{\Phi}_{g,c}(t,t_0)=A_{g,c}(t)\Phi_{g,c}(t,t_0)\qquad \Phi_{g,c}(t_0,t_0)=I_2
\]

If the carrier is rotor-synchronous and the operating speed is locally constant, then ${A}_{g,c}\left(t\right)$ is locally periodic. If the carrier is fixed-frequency or the speed varies over the window, the system should be treated as LTV.

The corresponding carrier-on output equation is

\[
{\tilde{\mathscr{y}}}_{c}\left(t\right)={C}_{g,c}\left(t\right){x}_{g}\left(t\right)+\nu \left(t\right)
\]

where $\nu \left(t\right)$ represents residual noise and modeling error. In the small-carrier expansion,

\[
{C}_{g,c}\left(t\right)={C}_{g,0}\left(t\right)+\epsilon {C}_{g,\epsilon }\left(t\right)+O({\epsilon}^{2})
\]

The passive degeneracy discussed in Section~\ref{sec:passive-coupling-degeneracy} corresponds to

\[
{C}_{g,0}\left(t\right)\approx 0
\]

for the selected deformation mode. The carrier-induced term is modeled as

\[
{C}_{g,\epsilon }\left(t\right)=c(t){\Gamma}_{\mathscr{y}}(m)\begin{bmatrix}1 & 0\end{bmatrix}
\]

Thus, under the output-dominant small-carrier approximation,

\[
{\tilde{\mathscr{y}}}_{c}\left(t\right)\approx \epsilon c(t){\Gamma}_{\mathscr{y}}\left(m\right)g(t)+\nu \left(t\right)
\]

Under the passive-degenerate output-dominant approximation, this corresponds to

\[
{C}_{g,c}\left(t\right)=\epsilon c(t){\Gamma}_{\mathscr{y}}(m)\begin{bmatrix}1 & 0\end{bmatrix}
\]

The carrier-induced observability Gramian for the air-gap state is

\[
\begin{aligned}W_{o,g}^{c}(t_0,t_f)={}&\int_{t_0}^{t_f}\Phi_{g,c}^{T}(t,t_0)C_{g,c}^{T}(t)R_c^{-1} \\&\quad{}C_{g,c}(t)\Phi_{g,c}(t,t_0)\,dt\end{aligned}
\]

where ${R}_{c}$ is the effective covariance of the carrier-synchronous residual channel.

Substituting the small-carrier output matrix gives

\[
\begin{aligned}W_{o,g}^{c}\approx{}&\epsilon^2\int_{t_0}^{t_f}c^2(t)\Phi_{g,c}^{T}(t,t_0)\begin{bmatrix}1\\0\end{bmatrix} \\&\quad{}\Gamma_{\mathscr{y}}^{T}(m)R_c^{-1}\Gamma_{\mathscr{y}}(m)\begin{bmatrix}1&0\end{bmatrix}\Phi_{g,c}(t,t_0)\,dt\end{aligned}
\]

Since

\[
{\Gamma}_{\mathscr{y}}^{T}(m){R}_{c}^{-1}{\Gamma}_{\mathscr{y}}(m)
\]

is scalar, this can be written as

\[
\begin{aligned}W_{o,g}^{c}\approx{}&\epsilon^2\left(\Gamma_{\mathscr{y}}^{T}(m)R_c^{-1}\Gamma_{\mathscr{y}}(m)\right) \\&\quad{}\int_{t_0}^{t_f}c^2(t)\Phi_{g,c}^{T}(t,t_0)\begin{bmatrix}1\\0\end{bmatrix}\begin{bmatrix}1&0\end{bmatrix}\Phi_{g,c}(t,t_0)\,dt\end{aligned}
\]

This expression shows that carrier-induced observability requires

\[
{\Gamma}_{\mathscr{y}}^{T}\left(m\right){R}_{c}^{-1}{\Gamma}_{\mathscr{y}}\left(m\right)>0
\]

If

\[
{\Gamma}_{\mathscr{y}}\left(m\right)=0
\]

then the carrier does not create an observation channel, regardless of its amplitude.

For analytical interpretation, one may further adopt an output-dominant approximation in which the carrier changes the residual channel more strongly than it changes the modal dynamics:

\[
{A}_{g,c}\left(t\right)\approx {A}_{g}(t)\qquad {\Phi}_{g,c}\left(t,{t}_{0}\right)\approx {\Phi}_{g}\left(t,{t}_{0}\right)
\]

Then the reduced expression becomes

\[
\begin{aligned}W_{o,g}^{c}\approx{}&\epsilon^2\left(\Gamma_{\mathscr{y}}^{T}(m)R_c^{-1}\Gamma_{\mathscr{y}}(m)\right) \\&\quad{}\int_{t_0}^{t_f}c^2(t)\Phi_g^{T}(t,t_0)\begin{bmatrix}1\\0\end{bmatrix}\begin{bmatrix}1&0\end{bmatrix}\Phi_g(t,t_0)\,dt\end{aligned}
\]

This reduced form is not a replacement for the full carrier-on LTV/LTP Gramian. It is a first-order analytical approximation that exposes the carrier-induced measurement mechanism. In full numerical evaluation, ${A}_{g,c}\left(t\right)$, ${C}_{g,c}\left(t\right)$, and ${\Phi}_{g,c}$ should be computed using the carrier-on dynamics.

When

\[
{\Gamma}_{\mathscr{y}}\left(m\right)\neq 0
\]

and the carrier has nonzero energy over the observation window, the residual provides a finite-time measurement of the deformation coordinate $g$. Through the air-gap dynamics,

\[
\dot{g}={v}_{g}
\]

finite-time evolution of this measurement can provide observability of both $g$ and ${v}_{g}$, subject to conditioning, bandwidth, and noise limitations.

\subsection{Sideband Detection in Production Residuals}
\label{subsec:sideband-detection}

The Gramian formulation establishes the carrier-induced observability mechanism. Sideband detection provides an implementation method.

Suppose the selected deformation is rotor-synchronous,

\[
g\left(t\right)={A}_{g}\cos(r{\theta}_{r}(t)+{\phi}_{g})
\]

and the injected carrier is

\[
{c}_{h}\left(t\right)=\cos(h{\theta}_{r}(t)+{\phi}_{h})
\]

the mixed term is

\[
g\left(t\right){c}_{h}\left(t\right)={A}_{g}\cos(r{\theta}_{r}+{\phi}_{g})\cos(h{\theta}_{r}+{\phi}_{h})
\]

Using the product identity,

\[
\begin{aligned}g(t)c_h(t)=\frac{A_g}{2}\Bigl[{}&\cos\!\left((h+r)\theta_r+\phi_h+\phi_g\right) \\&+\cos\!\left((h-r)\theta_r+\phi_h-\phi_g\right)\Bigr]\end{aligned}
\]

Thus, the selected deformation does not need to appear directly at $rX$. Under carrier-assisted observation, it may appear as sidebands at

\[
\left(h\pm r\right)X
\]

In frequency-domain form, for a time-domain carrier frequency ${\omega}_{c}$, the sidebands appear at

\[
{\omega}_{c}\pm r\Omega
\]

A lock-in detector can be applied to the carrier-synchronous residual:

\[
z_{h\pm r}=\int_{t_0}^{t_f}\tilde{\mathscr{y}}_c(t)e^{-j(h\pm r)\theta_r(t)}\,dt
\]

If carrier-induced observation exists, then the sideband magnitude scales approximately as

\[
\lvert z_{h\pm r}\rvert\propto\epsilon A_g\lvert\Gamma_{\mathscr{y}}(m)\rvert
\]

This scaling provides an experimental consistency check. A useful carrier should produce sidebands that are absent or much weaker when the carrier is off, and whose magnitude increases approximately linearly with carrier amplitude $\epsilon$ in the small-signal regime. Order-domain signal-processing methods are commonly used to separate speed-synchronous components and nearby orders in rotating machinery \cite{randall2011vibration,wang2009vold}.

Thus, the practical search for a useful observation carrier can be written as

\[
m,h,\phi_h,\omega_c,\epsilon\to\lvert z_{h+r}\rvert^2+\lvert z_{h-r}\rvert^2
\]

The lock-in detector is therefore an implementation of the carrier-induced output equation, not a substitute for the observability analysis. The observability condition remains governed by the carrier-dependent projection ${\Gamma}_{\mathscr{y}}\left(m\right)$ and the finite-time Gramian ${W}_{o,g}^{c}$.

The sideband signature should be interpreted as a controlled consistency check rather than a standalone proof of modal observability. Similar sidebands may also arise from inverter nonlinearities, PWM effects, slotting harmonics, load ripple, or unrelated electromagnetic asymmetries. Therefore, a useful validation should compare carrier-off and carrier-on residuals, sweep carrier amplitude and phase, and verify the expected small-signal scaling

\[
\lvert z_{h\pm r}\rvert\propto\epsilon A_g\lvert\Gamma_{\mathscr{y}}(m)\rvert
\]

Only sidebands that are repeatable, carrier-synchronous, phase-consistent, and distinguishable from the carrier-only baseline should be interpreted as evidence of carrier-induced observation.

\subsection{Carrier-Induced Force Projection and Reachability Gramian}
\label{subsec:carrier-force-reachability}

Carrier-induced observability does not automatically imply carrier-induced reachability. Observation depends on whether the selected deformation modulates a measurable electrical residual. Reachability depends on whether the injected carrier can generate generalized electromagnetic force along the selected deformation mode.

The generalized force associated with the air-gap deformation coordinate is

\[
{Q}_{g}\left(t\right)={\int}_{0}^{2\pi }{p}_{em}(\beta,t){\phi}_{r}(\beta)d\beta
\]

where ${p}_{em}$ is the electromagnetic pressure distribution. If the injected carrier creates an $r$-order pressure component,

\[
p_{em}(\beta,t)\supset P_r(t)\phi_r(\beta)
\]

then

\[
{Q}_{g}\left(t\right)\neq 0
\]

and the selected deformation mode becomes reachable through carrier-induced electromagnetic forcing.

Using the same field-mixing argument, the cross term between the main $p$-order field and a carrier $m$-order field generates pressure components at

\[
\lvert m-p\rvert=r
\]

Therefore, a carrier satisfying $m=p\pm r$ can, in the simplified field-mixing picture, produce an $r$-order electromagnetic pressure component. If this component survives the winding, inverter, saturation, and magnetic-map projections, it may have nonzero projection onto ${\phi}_{r}$, creating a generalized force path.

Define the carrier-induced force projection coefficient

\[
{\Gamma}_{F}\left(m\right)={\int}_{0}^{2\pi }{p}_{m}(\beta){\phi}_{r}(\beta)d\beta
\]

where ${p}_{m}(\beta)$ is the carrier-induced pressure shape. The active reachability condition is

\[
{\Gamma}_{F}\left(m\right)\neq 0
\]

With the carrier on, the air-gap subsystem is generally LTV or LTP:

\[
{\dot{x}}_{g}\left(t\right)={A}_{g,c}\left(t\right){x}_{g}\left(t\right)+{B}_{g,c}\left(t\right){u}_{c}\left(t\right)
\]

where ${u}_{c}\left(t\right)$ denotes the carrier-amplitude or carrier-phase control input associated with the selected carrier.

Since the generalized force enters the air-gap acceleration equation, the carrier-induced input matrix has the form

\[
{B}_{g,c}\left(t\right)=\frac{\epsilon {\Gamma}_{F}(m)}{{m}_{g}}c(t)\begin{bmatrix}0 \\ 1\end{bmatrix}
\]

The carrier-induced reachability Gramian is therefore

\[
\begin{aligned}W_{r,g}^{c}(t_0,t_f)={}&\int_{t_0}^{t_f}\Phi_{g,c}(t_f,\tau)B_{g,c}(\tau)R_{u,c}^{-1} \\&\quad{}B_{g,c}^{T}(\tau)\Phi_{g,c}^{T}(t_f,\tau)\,d\tau\end{aligned}
\]

where ${R}_{u,c}$ is the input weighting associated with the carrier command.

For a scalar carrier input, this becomes

\[
\begin{aligned}W_{r,g}^{c}\approx{}&\frac{\epsilon^2\Gamma_F^2(m)}{m_g^2}R_{u,c}^{-1}\int_{t_0}^{t_f}c^2(\tau)\Phi_{g,c}(t_f,\tau)\begin{bmatrix}0\\1\end{bmatrix} \\&\quad{}\begin{bmatrix}0&1\end{bmatrix}\Phi_{g,c}^{T}(t_f,\tau)\,d\tau\end{aligned}
\]

This expression shows that carrier-induced reachability requires

\[
{\Gamma}_{F}\left(m\right)\neq 0
\]

and nonzero carrier energy over the control horizon,

\[
{\int}_{{t}_{0}}^{{t}_{f}}{c}^{2}\left(\tau \right)d\tau >0
\]

For analytical interpretation, one may again use an output/input-dominant approximation,

\[
{A}_{g,c}\left(t\right)\approx {A}_{g}(t)\qquad {\Phi}_{g,c}\left({t}_{f},\tau \right)\approx {\Phi}_{g}\left({t}_{f},\tau \right)
\]

Then

\[
\begin{aligned}W_{r,g}^{c}\approx{}&\frac{\epsilon^2\Gamma_F^2(m)}{m_g^2}R_{u,c}^{-1}\int_{t_0}^{t_f}c^2(\tau)\Phi_g(t_f,\tau)\begin{bmatrix}0\\1\end{bmatrix} \\&\quad{}\begin{bmatrix}0&1\end{bmatrix}\Phi_g^{T}(t_f,\tau)\,d\tau\end{aligned}
\]

For a short-horizon approximation, let

\[
\Delta ={t}_{f}-\tau
\]

Neglecting damping and stiffness over an infinitesimal horizon,

\[
\Phi_g(t_f,\tau)\begin{bmatrix}0\\1\end{bmatrix}\approx\begin{bmatrix}\Delta\\1\end{bmatrix}
\]

Therefore,

\[
\begin{aligned}W_{r,g}^{c}\approx{}&\frac{\epsilon^2\Gamma_F^2(m)}{m_g^2}R_{u,c}^{-1}\int_{0}^{H}c^2(t_f-\Delta) \\&\quad{}\begin{bmatrix}\Delta^2&\Delta\\\Delta&1\end{bmatrix}\,d\Delta\end{aligned}
\]

This reduced expression is the carrier-induced analogue of the passive reachability Gramian. It shows explicitly that the carrier force projection enters through modal acceleration, not directly through modal displacement.

Carrier-induced reachability is a channel-existence and conditioning measure, not a guarantee of practical active suppression. Harmonic current injection and torque-ripple mitigation studies demonstrate that injected electrical harmonics can influence electromagnetic torque and force components, but practical implementation remains constrained by inverter, current-loop, loss, and NVH limitations \cite{jahns1996pulsating,yan2019torque,erken2016online,dai2026data,soresini2025numerical}. Practical force authority depends on whether the inverter and current regulator can synthesize the required carrier without violating DC-bus voltage limits, current limits, PWM resolution, carrier-frequency ratio, field-weakening margin, thermal and loss budgets, acoustic constraints, or electromagnetic compatibility constraints. A carrier may satisfy ${\Gamma}_{F}\left(m\right)\neq 0$ and still be unusable if the required amplitude is too large or if it introduces unacceptable torque ripple, copper loss, iron loss, or additional NVH components.

Therefore, the reachability condition in this section should be interpreted as a necessary feasibility condition for active excitation or suppression, not as a sufficient condition for closed-loop attenuation.

\subsection{Order-Domain Verification of Carrier-Induced Reachability}
\label{subsec:order-domain-reachability-verification}

The reachability Gramian in Section~\ref{subsec:carrier-force-reachability} establishes whether the injected carrier can generate generalized electromagnetic force along the selected air-gap deformation mode. In implementation, this force path is usually not measured directly. It must be inferred from the response of the system to a known carrier command.

Let the carrier command be

\[
{u}_{c}\left(t\right)={U}_{c}\cos(h{\theta}_{r}(t)+{\phi}_{u})
\]

where ${U}_{c}$ and ${\phi}_{u}$ are the carrier amplitude and phase. If the selected carrier produces a nonzero force projection, 
${\Gamma}_{F}\left(m\right)\neq 0$ then the air-gap subsystem receives a carrier-induced modal forcing term,

\[
{\dot{x}}_{g}\left(t\right)={A}_{g,c}\left(t\right){x}_{g}\left(t\right)+{B}_{g,c}\left(t\right){u}_{c}\left(t\right)
\]

The resulting deformation response is given by the finite-time input-to-state map

\[
\begin{aligned}x_g(t)={}&\Phi_{g,c}(t,t_0)x_g(t_0) \\&+\int_{t_0}^{t}\Phi_{g,c}(t,\tau)B_{g,c}(\tau)u_c(\tau)\,d\tau\end{aligned}
\]

If a carrier-dependent observation channel is also available,

\[
{\Gamma}_{\mathscr{y}}\left(m\right)\neq 0
\]

then the modal response appears in the carrier-synchronous residual,

\[
{\tilde{\mathscr{y}}}_{c}\left(t\right)={C}_{g,c}\left(t\right){x}_{g}\left(t\right)+\nu \left(t\right)
\]

Substituting the input-to-state response gives the production-measurable input-output relation

\[
\begin{aligned}\tilde{\mathscr{y}}_c(t)\approx{}&C_{g,c}(t)\int_{t_0}^{t}\Phi_{g,c}(t,\tau)B_{g,c}(\tau)u_c(\tau)\,d\tau \\&+\nu(t)\end{aligned}
\]

after neglecting the free response or after subtracting the carrier-off baseline.

Thus, a nonzero production-measurable reachability signature requires both a force path and a measurement path:

\[
{\Gamma}_{F}\left(m\right)\neq 0\qquad {\Gamma}_{\mathscr{y}}\left(m\right)\neq 0
\]

Carrier-induced reachability can be checked by sweeping carrier amplitude and phase and observing the carrier-synchronous residual response. In the small-signal regime, the response magnitude should scale approximately with carrier amplitude,

\[
\lvert z_{\mathscr{y}}\rvert\propto U_c\lvert\Gamma_F(m)\rvert\lvert\Gamma_{\mathscr{y}}(m)\rvert\lvert\mathscr{H}_g\rvert
\]

where ${\mathscr{H}}_{g}$ denotes the finite-time modal compliance from carrier-induced generalized force to the air-gap deformation coordinate.

This production-residual test verifies the combined input-output path, not the force path alone. If ${\Gamma}_{\mathscr{y}}\left(m\right)$ is weak or zero, a nonzero ${\Gamma}_{F}\left(m\right)$ may still exist but remain invisible in production electrical residuals. In that case, carrier-induced force authority must be validated using external NVH or vibration measurements, such as housing acceleration, acoustic response, shaft vibration, or dynamometer instrumentation.

A phase sweep provides an additional consistency check. If the carrier generates modal force, changing the carrier phase ${\phi}_{u}$ should rotate the phase of the measured residual response. Near resonance, the response phase should also reflect the dynamics of the air-gap subsystem.

For order-domain implementation, the response can be evaluated using the same carrier sideband structure discussed in Section~\ref{subsec:sideband-detection}. A carrier command at order $h$ interacting with a selected deformation order $r$ produces measurable residual components at

\[
\left(h\pm r\right)X
\]

provided that the carrier both excites the mode and allows the resulting response to be observed. Therefore, the reachability verification metric can be written as

\[
J_F(m)=\lvert z_{h+r}\rvert^2+\lvert z_{h-r}\rvert^2
\]

evaluated under controlled sweeps of ${U}_{c}$, ${\phi}_{u}$, and operating condition.

If no production-accessible observation channel exists, production data alone cannot verify reachability. In that case, a nonzero ${\Gamma}_{F}\left(m\right)$ must be validated using external NVH or vibration measurements, such as housing acceleration, acoustic response, shaft vibration, or dynamometer instrumentation. This distinction is important: carrier-induced force authority may exist even when the production electrical residual remains weak.

\subsection{Summary and Scope of Carrier-Induced Coupling}
\label{subsec:carrier-coupling-summary}

The carrier-induced framework changes the passive projection problem. In the passive model, a selected low-order air-gap deformation may satisfy ${\ell}_{\lambda }\approx 0$, causing the first-order current-observation and voltage-reachability paths to collapse. With an injected carrier, the relevant quantities become the carrier-dependent measurement and force projections, ${\Gamma}_{\mathscr{y}}\left(m\right)$ and ${\Gamma}_{F}\left(m\right)$.

The condition

\[
{\Gamma}_{\mathscr{y}}\left(m\right)\neq 0
\]

indicates that the selected deformation can modulate a production-accessible residual channel. The condition

\[
{\Gamma}_{F}\left(m\right)\neq 0
\]

indicates that the carrier can generate generalized electromagnetic force along the selected deformation mode.

These conditions are related but not equivalent. Diagnostic observation requires ${\Gamma}_{\mathscr{y}}\left(m\right)\neq 0$, active excitation or suppression requires ${\Gamma}_{F}\left(m\right)\neq 0$, and closed-loop active attenuation requires both

\[
{\Gamma}_{\mathscr{y}}\left(m\right)\neq 0\qquad {\Gamma}_{F}\left(m\right)\neq 0
\]

The useful carrier is not arbitrary. For a selected low-order deformation of order $r$, the simplified field-mixing condition is

\[
m=p\pm r
\]

with

\[
m\in {\mathscr{M}}_{act}
\]

Thus, the feasibility of carrier-induced observability or reachability depends on whether the winding, inverter, current regulator, magnetic saturation, and allowable harmonic injection can synthesize a carrier that creates a nonzero projection for the selected mode. Carrier-induced probing is therefore not ordinary small-signal probing through the passive linear channel. It introduces a known electromagnetic carrier that can create a new bilinear measurement or force projection.

Hence

\[
{\ell}_{\lambda }\approx 0
\]

does not imply

\[
\Gamma_{\mathscr{y}}(m)=0\qquad\text{or}\qquad\Gamma_F(m)=0
\]

This reframes the limitation from passive observability to active carrier design.

The coefficients ${\Gamma}_{\mathscr{y}}\left(m\right)$ and ${\Gamma}_{F}\left(m\right)$ should be interpreted as carrier-on feasibility quantities. In a production validation workflow, they should be estimated from FEM-calibrated co-energy maps, carrier-on electromagnetic simulation, controlled carrier sweeps, or residual measurements rather than assumed from the simplified harmonic closure alone. The simplified field-mixing argument identifies promising carrier orders and projection mechanisms; it does not replace numerical magnetic calibration, controller modeling, or experimental validation.

\section{Physically Anchored Numerical Feasibility Studies}
\label{sec:numerical-feasibility}

The previous sections derived the passive and carrier-induced coupling mechanisms using magnetic co-energy, finite-time Gramians, and local projection arguments. This section provides two numerical feasibility studies that make those mechanisms computationally explicit. The objective is not to validate the production motor or predict absolute current, voltage, vibration, or acoustic amplitudes. Instead, the objective is to test two framework-level questions under physically anchored but non-FEM-calibrated assumptions.

First, the passive projection study evaluates whether selected low-order deformation components survive a symmetry-preserving dq projection when the motor architecture is represented by a 48-slot, 8-pole configuration. Second, the carrier-induced residual study evaluates how large the composite carrier-induced projection must be before a sideband becomes detectable under finite-window residual noise and near-order disturbance leakage.

These studies should therefore be interpreted as numerical feasibility demonstrations of the proposed framework. They do not replace FEM-calibrated magnetic maps, structural modal validation, controller-in-the-loop simulation, inverter-constrained carrier synthesis, or dynamometer experiments. High-fidelity PMSM NVH prediction generally requires coupled electromagnetic, structural, and acoustic modeling or FEM-calibrated force maps \cite{soresini2024noise,wang2023modelling,yang2020radial,muller2025evaluation}.

\subsection{Architecture-Aware Passive Projection Study}
\label{subsec:architecture-aware-passive-projection}

The first study evaluates the passive projection mechanism derived in Section~\ref{sec:passive-coupling-degeneracy}. The motivating machine architecture is represented as a 48-slot, 8-pole electric machine. Therefore, the total number of magnetic poles and the pole-pair number are

\[
N_{\mathrm{pole}}=8\qquad p=\frac{N_{\mathrm{pole}}}{2}=4
\]

The ideal continuous dq basis is defined as

\[
b_d(\beta)=\cos(p\beta)\qquad b_q(\beta)=\sin(p\beta)
\]

where $\beta$ is the rotor-fixed circumferential coordinate. The selected deformation component is represented as

\[
\phi_r(\beta)=\cos(r\beta+\phi_g)
\]

where $r$ is the selected low-order deformation order and $\phi_g$ is the deformation phase.

To evaluate the projection numerically, a representative normalized PM MMF field is used,

\[
\mathscr{F}_m(\beta)=\sum_{n\in\mathscr{H}_m}a_n\cos(np\beta+\varphi_n)
\]

where the baseline study includes the fundamental and selected higher PM harmonics. The purpose of this field representation is not to reproduce the real magnetic field. It only provides a normalized electromagnetic kernel through which the low-order deformation is projected.

For each selected deformation order, the PM flux-linkage sensitivity terms are approximated by the normalized spatial averages

\[
\psi_{d,g}^{(r)}=-\left\langle M(\beta)b_d(\beta)\mathscr{F}_m(\beta)\phi_r(\beta)\right\rangle
\]

\[
\psi_{q,g}^{(r)}=-\left\langle M(\beta)b_q(\beta)\mathscr{F}_m(\beta)\phi_r(\beta)\right\rangle
\]

where $\left\langle\cdot\right\rangle$ denotes the full-circumference average and $M(\beta)$ is the projection mask. The ideal continuous case uses

\[
M(\beta)=1
\]

A second architecture-aware proxy is also evaluated using a 48-slot winding kernel with six skew layers. This proxy preserves the slot count, pole-pair number, distributed winding structure, and axial skew averaging, but it is not a measured winding function or FEM-derived magnetic map.

The inductance-sensitivity terms are approximated similarly:

\[
L_{dd,g}^{(r)}=-\left\langle M(\beta)b_d^2(\beta)\phi_r(\beta)\right\rangle
\]

\[
L_{dq,g}^{(r)}=-\left\langle M(\beta)b_d(\beta)b_q(\beta)\phi_r(\beta)\right\rangle
\]

\[
L_{qq,g}^{(r)}=-\left\langle M(\beta)b_q^2(\beta)\phi_r(\beta)\right\rangle
\]

For a representative normalized loaded operating point $(I_{d0},I_{q0})$, the passive flux-linkage sensitivity vector is then computed as

\[
\ell_\lambda^{(r)}=\begin{bmatrix}\lambda_{d,g}^{(r)}\\\lambda_{q,g}^{(r)}\end{bmatrix}
\]

where

\[
\lambda_{d,g}^{(r)}=\psi_{d,g}^{(r)}+I_{d0}L_{dd,g}^{(r)}+I_{q0}L_{dq,g}^{(r)}
\]

\[
\lambda_{q,g}^{(r)}=\psi_{q,g}^{(r)}+I_{d0}L_{dq,g}^{(r)}+I_{q0}L_{qq,g}^{(r)}
\]

The reported magnitude is normalized by the projection obtained from a uniform air-gap perturbation, $r=0$. Therefore,

\[
\left\|\ell_\lambda(r)\right\|_{norm}=\frac{\max_{\phi_g}\left\|\ell_\lambda^{(r)}(\phi_g)\right\|_2}{\max_{\phi_g}\left\|\ell_\lambda^{(0)}(\phi_g)\right\|_2}
\]

This normalization avoids assigning physical units to the non-FEM-calibrated model and allows the study to focus on relative projection strength.

\begin{figure}[t]
\centering
\includegraphics[width=\columnwidth]{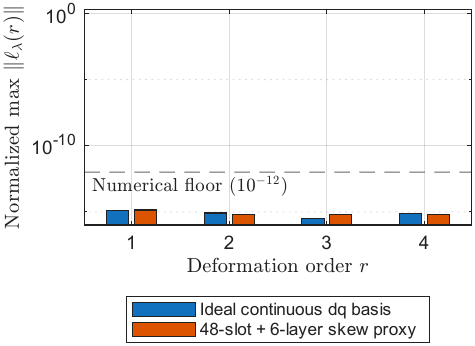}
\caption{Normalized passive projection magnitude for selected low-order deformation orders under an ideal continuous dq basis and a 48-slot six-layer skew proxy.}
\label{fig:passive-projection}
\end{figure}

Fig.~\ref{fig:passive-projection} shows that the selected low-order deformation orders

\[
r=1,2,3,4
\]

remain below the numerical floor under both the ideal continuous dq basis and the 48-slot six-layer skew proxy. This indicates that the architecture-aware proxy preserves the same symmetry-conditioned cancellation predicted by the ideal continuous projection model. Under this symmetric closure, the selected low-order deformation can be mechanically present while producing no first-order passive dq electrical projection.

To evaluate whether this passive cancellation is structurally fragile, a controlled low-order symmetry-breaking perturbation is introduced:

\[
M_\eta(\beta)=1+\eta\cos(2\beta+\phi_\eta)
\]

where $\eta$ is the low-order asymmetry amplitude. This perturbation is not intended to represent a specific manufacturing defect. It is used only to test how an order-2 symmetry-breaking component can reopen the passive projection path.

\begin{figure}[t]
\centering
\includegraphics[width=\columnwidth]{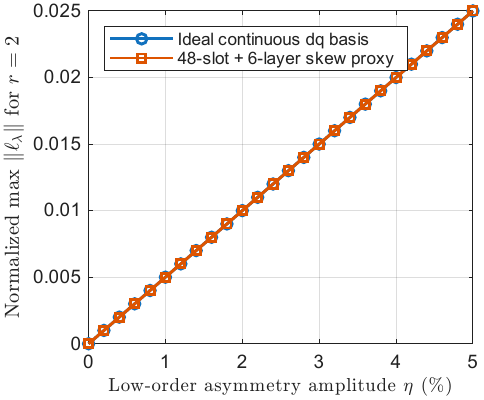}
\caption{Reopening of the $r=2$ passive projection under a controlled low-order symmetry-breaking perturbation.}
\label{fig:asymmetry-reopening}
\end{figure}

Fig.~\ref{fig:asymmetry-reopening} shows that the normalized passive projection for the motivating $r=2$ case increases approximately linearly with $\eta$. This result demonstrates that passive invisibility is a symmetry-conditioned cancellation result, not a universal absence of electromechanical coupling. In a real machine, manufacturing variation, rotor or stator eccentricity, localized saturation, magnet-to-magnet variation, skew imperfections, or calibrated magnetic-map asymmetry may reopen a weak passive electrical path.

\subsection{Carrier-Induced Residual Detectability Study}
\label{subsec:carrier-residual-detectability}

The second study evaluates the detectability of a carrier-induced residual sideband under finite-window noise and near-order disturbance conditions. This study supports the carrier-induced observability mechanism developed in Section~\ref{sec:carrier-induced-projection}.

The carrier-on residual model is written in order-domain form as

\[
\tilde{\mathscr{y}}_c(\theta)=\alpha\cos(h\theta+\phi_h)\cos(r\theta+\phi_g)+\nu(\theta)
\]

where $\theta$ is the mechanical angle, $r$ is the selected deformation order, $h$ is the carrier order, and $\nu(\theta)$ represents residual noise. The composite projection strength is

\[
\alpha=\epsilon A_g\left\|\Gamma_{\mathscr{y}}\right\|
\]

where $\epsilon$ is the carrier amplitude, $A_g$ is the deformation amplitude, and $\left\|\Gamma_{\mathscr{y}}\right\|$ is the carrier-induced measurement projection magnitude.

The study does not assume a calibrated value of $\Gamma_{\mathscr{y}}$. Instead, $\alpha$ is treated as an independent variable. This reframes carrier-induced observability as a detectability-threshold problem: How large must $\alpha$ be for the carrier-induced sideband to exceed the residual noise floor?

For the motivating Order-2 case,

\[
r=2
\]

The carrier order is selected as

\[
h=6
\]

which corresponds to the $p+r$ candidate for $p=4$. The target sidebands are therefore

\[
h-r=4X\qquad h+r=8X
\]

Using the product identity

\[
\begin{aligned}\cos(h\theta+\phi_h)\cos(r\theta+\phi_g)={}&\frac{1}{2}\cos\!\left((h+r)\theta+\phi_h+\phi_g\right)\\&+\frac{1}{2}\cos\!\left((h-r)\theta+\phi_h-\phi_g\right)\end{aligned}
\]

each ideal sideband has amplitude

\[
\frac{\alpha}{2}
\]

A lock-in detector is used to estimate the sideband amplitude at a selected order $s$:

\[
z_s=2\left|\frac{\sum_k\mathscr{w}_k\tilde{\mathscr{y}}_c(\theta_k)e^{-js\theta_k}}{\sum_k\mathscr{w}_k}\right|
\]

where $\mathscr{w}_k$ is a Hann window over the finite observation interval. The sideband response is evaluated at both

\[
s=h\pm r
\]

The sideband SNR is defined as

\[
\mathrm{SNR}=\frac{\sqrt{\frac{1}{2}\left(z_{h-r}^2+z_{h+r}^2\right)}}{\sigma_z}
\]

where $\sigma_z$ is the lock-in noise floor estimated from noise-only Monte Carlo trials.

\begin{figure}[t]
\centering
\includegraphics[width=\columnwidth]{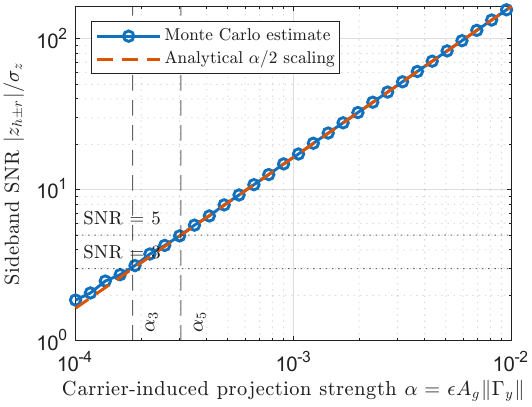}
\caption{Sideband SNR as a function of the composite carrier-induced projection strength $\alpha=\epsilon A_g\lVert\Gamma_{\mathscr{y}}\rVert$}
\label{fig:sideband-snr}
\end{figure}

Fig.~\ref{fig:sideband-snr} shows the sideband SNR as a function of $\alpha$. The Monte Carlo estimate follows the analytical $\alpha/2$ sideband scaling, confirming that the lock-in detector recovers the carrier-induced residual projection in the finite-window simulation. More importantly, the plot identifies the projection strengths

\[
\alpha_3\qquad \alpha_5
\]

which indicate the projection strength required to exceed SNR levels of 3 and 5. Since the ideal sideband amplitude is $\alpha/2$, these thresholds satisfy approximately

\[
\alpha_3=6\sigma_z\qquad \alpha_5=10\sigma_z
\]

Thus, a nonzero carrier-induced projection coefficient is not sufficient by itself. Practical residual observability requires the composite projection strength

\[
\epsilon A_g\left\|\Gamma_{\mathscr{y}}\right\|
\]

to exceed a finite-window noise threshold.

The second part of the study evaluates false lock-in caused by near-order disturbances. The target Order-2 component is turned off, and near-order disturbance components are introduced at

\[
1.85X\qquad 1.95X\qquad 2.05X\qquad 2.15X
\]

These components are mixed with the same carrier order $h=6$. Although they do not generate sidebands exactly at $4X$ and $8X$, finite-window spectral leakage can cause them to appear in the $h\pm r$ lock-in detectors.

\begin{figure}[t]
\centering
\includegraphics[width=\columnwidth]{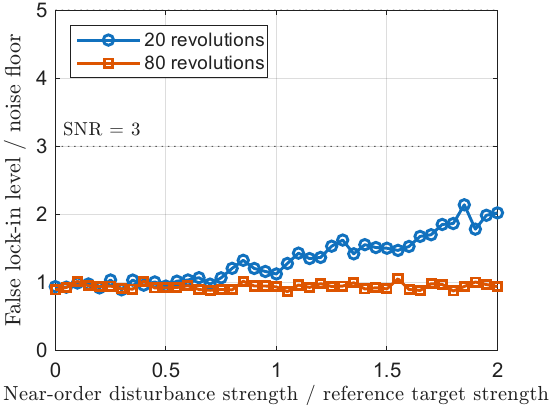}
\caption{False lock-in response at the target $h\pm r$ sidebands caused by near-order disturbances at $1.85X$, $1.95X$, $2.05X$, and $2.15X$.}
\label{fig:near-order-leakage}
\end{figure}

Fig.~\ref{fig:near-order-leakage} compares false lock-in levels for 20-revolution and 80-revolution observation windows. The short 20-revolution window shows increasing false lock-in response as the near-order disturbance strength increases. In contrast, the 80-revolution window remains close to the noise floor over the tested range. This indicates that carrier-synchronous residual validation must account for observation-window length and near-order disturbance leakage. Longer order-domain observation windows improve discrimination between true $h\pm r$ carrier-induced sidebands and nearby disturbance-induced components. The use of longer order-domain observation windows to improve separation of nearby components is consistent with rotating-machinery order-domain signal-processing practice \cite{randall2011vibration,wang2009vold}.

\subsection{Interpretation and Limitations}
\label{subsec:numerical-interpretation-limitations}

The two numerical studies support the projection and detectability mechanisms developed in the previous sections, but they do not validate the real production motor.

The first study shows that the passive-degeneracy mechanism survives a symmetry-preserving architecture-aware projection proxy. Under both the ideal continuous dq basis and the 48-slot six-layer skew proxy, the selected low-order deformation components remain below the numerical floor. This supports the argument that a mechanically present low-order mode may be first-order invisible to passive dq electrical projection under symmetric conditions. The symmetry-breaking sweep further shows that this invisibility can be weakened by low-order asymmetry.

The second study shows that carrier-induced observability should be interpreted as a finite-window residual detectability problem. The carrier-induced projection must be strong enough to exceed the residual noise floor, and the detected sideband must be distinguishable from near-order disturbance leakage. Therefore, a sideband signature should not be treated as standalone proof of modal observability. It should be interpreted together with carrier-off baselines, carrier-on repeatability, phase consistency, observation-window length, and disturbance rejection.

The limitations of these studies are explicit. The passive projection study does not use FEM-calibrated magnetic maps, measured winding functions, measured PM fields, or real manufacturing variations. The carrier-induced detectability study does not estimate the real value of $\Gamma_{\mathscr{y}}(m)$, does not model inverter voltage constraints, does not include current-loop dynamics, and does not use measured residual noise spectra. Therefore, the results should be read as numerical feasibility demonstrations of the framework, not as predictions of production motor behavior.

A complete validation workflow would require FEM-calibrated co-energy maps, structural or experimental modal identification of the selected air-gap deformation coordinate, carrier-on electromagnetic simulation, controller and inverter feasibility analysis, and controlled dynamometer experiments with production signal logging and external NVH reference measurements.

\section{Conclusions and Future Validation Roadmap}
\label{sec:conclusions-validation}

This paper developed a sensor-limited observability and carrier-induced reachability framework for low-order rotor-coupled NVH in production electric drives. The motivating problem is a low-order rotor-coupled vibration component that is mechanically relevant under load but not directly measured by production sensors. The central question is therefore whether such a mode can be inferred or influenced using production-accessible current, voltage-command, resolver, and carrier-synchronous residual signals.

The analysis represented the selected vibration component as a reduced air-gap deformation state,

\[
x_g=\begin{bmatrix}g\\v_g\end{bmatrix}
\]

and embedded it into a magnetic co-energy-based electromechanical model. Finite-time Gramians were then used to evaluate passive current observability, resolver observability, and plant-level voltage-to-air-gap reachability. The passive analysis identified the flux-linkage sensitivity vector $\ell_\lambda$ as the common bottleneck for passive current observability and passive voltage reachability, while resolver observability was shown to depend on a separate torque-projection condition.

This paper further showed that passive dq invisibility does not necessarily imply carrier-induced invisibility. With an injected carrier, the relevant feasibility quantities become the carrier-dependent measurement and force projections, $\Gamma_{\mathscr{y}}(m)$ and $\Gamma_F(m)$. Two physically anchored numerical studies were included to demonstrate the computational structure of the framework: an architecture-aware passive projection study and a carrier-induced residual detectability study. These studies support the feasibility logic but do not replace FEM-calibrated magnetic maps, controller/inverter modeling, or experimental validation.

\subsection{Main Conclusions}
\label{subsec:main-conclusions}

The main conclusions are as follows.

First, a low-order rotor-coupled air-gap deformation can be mechanically relevant while remaining first-order invisible to passive dq electrical signals. This occurs when the selected deformation is orthogonal to the winding-level dq electromagnetic projection.

Second, passive current observability and passive voltage-to-air-gap reachability share the same electromagnetic bottleneck, $\ell_\lambda$. When $\ell_\lambda\approx0$, both the passive current-observation path and the passive plant-level voltage-to-air-gap force path become locally degenerate.

Third, resolver observability is governed by a separate torque-projection condition, $\rho_T$. A nonzero flux-linkage sensitivity vector does not automatically imply a usable resolver signature, because the deformation-induced torque must survive rotor inertia, filtering, estimator bandwidth, quantization, and noise.

Fourth, passive invisibility does not imply carrier-induced invisibility. A carrier can introduce a bilinear measurement or force projection through carrier-dependent coefficients $\Gamma_{\mathscr{y}}(m)$ and $\Gamma_F(m)$. Therefore,

\[
\ell_\lambda\approx0
\]

does not imply

\[
\Gamma_{\mathscr{y}}(m)=0\qquad \Gamma_F(m)=0
\]

Fifth, carrier-induced observability and reachability are distinct. A carrier may create a measurable residual projection without generating useful modal force, or it may generate modal force that remains invisible to production electrical residuals. Therefore, observation feasibility and force-authority feasibility must be evaluated separately.

Sixth, carrier-induced residual detection is a finite-window signal-conditioning problem. A nonzero $\Gamma_{\mathscr{y}}(m)$ is not sufficient by itself. The composite projection strength $\epsilon A_g\left\|\Gamma_{\mathscr{y}}\right\|$ must exceed the residual noise floor and remain distinguishable from carrier-off baselines, carrier-only response, near-order leakage, inverter nonlinearities, and unrelated harmonic disturbances.

\subsection{Future Validation Roadmap}
\label{subsec:future-validation-roadmap}

This paper establishes the theoretical and computational framework, but it does not provide full production-motor validation. A complete validation workflow should proceed through five high-level stages.

\subsubsection{FEM-Calibrated Magnetic Projection}
\label{subsubsec:fem-calibrated-magnetic-projection}

The first validation step is to replace the analytical co-energy closure with FEM-calibrated or experimentally identified magnetic maps. FEM-calibrated electromagnetic force, flux, and skew effects are commonly required for quantitative PMSM NVH prediction \cite{soresini2024noise,wang2023modelling,yang2020radial,muller2025evaluation}. The required maps are

\[
\mathscr{w}_c(I_d,I_q,\theta_r,g)
\]

\[
\lambda_d(I_d,I_q,\theta_r,g)\qquad \lambda_q(I_d,I_q,\theta_r,g)
\]

and their derivatives with respect to the selected air-gap deformation coordinate. The key quantities to estimate are

\[
\lambda_{d,g}\qquad \lambda_{q,g}\qquad \mathscr{w}_{c,\theta_r,g}\qquad \mathscr{w}_{c,gg}
\]

These quantities determine the passive current-observation path, the passive voltage-to-air-gap reachability path, the resolver torque-projection path, and the electromagnetic stiffness contribution.

For the carrier-on problem, FEM-calibrated or experimentally identified magnetic maps should also be used to estimate

\[
\Gamma_{\mathscr{y}}(m)\qquad \Gamma_F(m)
\]

rather than assuming them from the simplified harmonic field-mixing argument alone. This step would convert the current analytical feasibility framework into a calibrated electromagnetic projection model.

\subsubsection{Reduced Air-Gap Modal Validation}
\label{subsubsec:reduced-air-gap-modal-validation}

The second validation step is to identify whether the selected air-gap coordinate $g$ corresponds to a measurable structural or operational deformation component. This paper treats $g$ as a reduced electromagnetic projection coordinate, not as a full structural model. A validation workflow should therefore estimate or identify the modal parameters

\[
m_g\qquad c_g\qquad k_g
\]

and relate the selected deformation order $r$ to measured vibration, housing acceleration, shaft response, acoustic response, or operational deflection behavior.

This step may use structural FEM, rotor-bearing-stator modal analysis, impact or shaker testing, operational vibration measurements, or dynamometer-based modal response fitting. The objective is not to replace the electromagnetic projection framework, but to ground the reduced coordinate $g$ in a physically identifiable deformation or response component.

\subsubsection{Carrier-On LTV/LTP Simulation}
\label{subsubsec:carrier-on-ltv-ltp-simulation}

The third validation step is to simulate the carrier-on dynamics using calibrated electromagnetic and modal quantities. Once a carrier is injected, the relevant system is generally LTV, or LTP when the carrier is rotor-synchronous and the operating speed is locally constant. Therefore, future simulation should compute

\[
A_{g,c}(t)\qquad C_{g,c}(t)\qquad B_{g,c}(t)\qquad \Phi_{g,c}(t,t_0)
\]

and the corresponding carrier-on Gramians,

\[
W_{o,g}^{c}\qquad W_{r,g}^{c}
\]

This step should evaluate how carrier amplitude, carrier phase, carrier order, operating speed, current operating point, saturation level, and observation-window length affect carrier-induced observability and reachability. Floquet analysis, harmonic balance, or direct finite-time LTV integration may be used depending on the carrier implementation and operating condition.

\subsubsection{Controller and Inverter Feasibility}
\label{subsubsec:controller-inverter-feasibility}

The fourth validation step is to evaluate whether the carrier required by the projection analysis can be implemented by the production inverter and current controller. The carrier must be feasible not only electromagnetically, but also electrically and thermally.

The relevant constraints include DC-bus voltage, voltage headroom, field-weakening margin, current limit, current-loop bandwidth, command filtering, PWM frequency, inverter dead time, sampling rate, thermal and loss budgets, acoustic side effects, and electromagnetic compatibility constraints.

This stage should determine whether a carrier satisfying

\[
m=p\pm r\qquad m\in\mathscr{M}_{act}
\]

can be synthesized with acceptable voltage, current, loss, and NVH side effects. It should also determine whether the current regulator suppresses the desired perturbation, whether the carrier remains visible in voltage-equation residuals, and whether the injected carrier creates unacceptable torque ripple or new acoustic components.

\subsubsection{Experimental Validation with Production Signals and External NVH References}
\label{subsubsec:experimental-production-signals-nvh}

The final validation step is controlled experimental testing. Production-accessible logs should include

\[
I_d\qquad I_q\qquad V_d^{cmd}\qquad V_q^{cmd}\qquad\theta_r\qquad\omega_r
\]

while external reference measurements should include at least one independent NVH or vibration channel, such as housing acceleration, microphone response, laser vibrometer measurement, shaft vibration, torque ripple, or dynamometer instrumentation.

The experimental workflow should compare carrier-off residuals, carrier-on residuals, carrier-only baselines, and external NVH response. A useful carrier-induced observation signature should be repeatable, carrier-synchronous, phase-consistent, and distinguishable from near-order leakage and unrelated harmonic disturbances. A useful force-path signature should show that the injected carrier can influence the external NVH or vibration response associated with the selected mode.  External vibration and order-domain measurements are standard tools for rotating-machine validation and condition monitoring \cite{randall2011vibration,wang2009vold}.

Closed-loop attenuation should be treated as a later objective. The present framework establishes observability and reachability feasibility. Active suppression would require additional controller design, stability analysis, robustness assessment, constraint handling, and validation that the carrier does not introduce unacceptable secondary NVH or loss.

\subsection{Closing Statement}
\label{subsec:closing-statement}

The main contribution of this paper is a feasibility framework for deciding whether a production electric drive has enough signal and actuation structure to observe or influence a selected low-order rotor-coupled NVH mode. The framework separates four questions that are often conflated: passive electrical visibility, resolver torque-path visibility, carrier-induced residual observability, and carrier-induced force reachability.

This separation is important for production electric drives because a low-order vibration may be mechanically important yet nearly invisible to passive electrical measurements. Conversely, an active carrier may create a measurable residual projection or a force projection even when the passive dq projection is degenerate. The practical value of the approach is therefore not the assumption that carrier injection always works, but the ability to define when it can work, what must be calibrated, and what failure modes must be checked before attempting active NVH suppression.

The framework remains incomplete without calibrated magnetic maps, structural or experimental modal validation, controller and inverter feasibility analysis, and dynamometer testing. However, it provides a structured path from a production NVH observation to a research program in sensor-limited estimation, electromagnetic projection, and learning- or probing-assisted control of intelligent physical systems under limited measurement conditions.

\bibliographystyle{IEEEtran}
\bibliography{references}

\end{document}